\documentclass[aps,preprint]{revtex4-1}%
\usepackage{amssymb}
\usepackage{amsfonts}
\usepackage{amsmath}
\usepackage{graphicx}%
\providecommand{\U}[1]{\protect\rule{.1in}{.1in}}
\newtheorem{theorem}{Theorem}
\newtheorem{acknowledgement}[theorem]{Acknowledgement}

\begin{document}
\preprint{CTCMP-CNU: 5}
\title[On FQHE]{On Fractional Quantum Hall Effect (FQHE): A Chern-Simons and nonequilibrium
quantum transport Weyl transform approach}
\author{F.A Buot,$^{1,2,3}$ G. Maglasang,$^{1,3}$A.R.F. Elnar,$^{1,3}$ and C.M.
Galon$^{1}$}
\affiliation{$^{1}$Center for Theoretical Condensed Matter Physics (CTCMP), Cebu Normal
University, Cebu City 6000, Philippines,}
\affiliation{$^{2}$C\&LB Research Institute, Carmen 6005, Philippines, }
\affiliation{$^{3}$LCFMNN, University of San Carlos, Cebu City 6000, Philippines}
\pacs{PACS number}

\begin{abstract}
We give a simple macroscopic phase-space explanation of fractional quantum
Hall effect (FQHE), in a fashion reminiscent of the Landau-Ginsburg
macroscopic symmetry breaking analyses. This is in contrast to the more
complicated microscopic wavefunction approaches. Here, we employ a
nonequilibrium quantum transport in the lattice Weyl transform formalism. This
is coupled with the Maxwell Chern-Simons gauge theory for defining fractional
filling of Landau levels. Flux attachment concept is inherent in fully
occupied and as well as in partially occupied Landau levels. We derived the
$k$-factor scaling hierarchy in Chern-Simons gauge theory, as the scaling
hierarchy of the magnetic fields or magnetic flux in FQHE. This is crucial in
our simple explanation of FQHE as a topological invariant in phase space. For
the fundamental scaling hierarchy, the integer $k$ must be a prime number, and
for fractions both the numerator and denominator of $k$ must also be prime
numbers. The assumption in the literature that a hierarchy of denominators of
$v=\frac{1}{k}$ is given by the expression, $\left(  2n+1\right)  $, is wrong.
Furthermore, even denominators for $v$ cannot belong to fundamental scaling
hierarchy and is often absent or less resolved in the experiments.


\end{abstract}
\volumeyear{year}
\volumenumber{number}
\issuenumber{number}
\eid{identifier}
\date[August 27, 2021]{}
\startpage{1}
\endpage{ }
\maketitle

\section{INTRODUCTION}

In previous papers \cite{previous, comments, jagna, covar}, we make use of the
gapped energy-band structure of solids under external electric field to derive
the integer quantum Hall effect (IQHE) of Chern insulator. We employ the
real-time superfield and lattice Weyl transform nonequilibrium Green's
function (SFLWT-NEGF) \cite{buot8} quantum transport formalism \cite{buot1,
bj} to the first-order gradient expansion to derive the topological Chern
number of the IQHE for two-dimensional systems, as an integral multiple of
quantum conductance, also known as the minimal contact conductance in
mesoscopic physics \cite{buot8}.

We find that the quantization of Hall effect occurs strictly not to first
order in the electric field \textit{per se} but rather to first-order gradient
expansion in the nonequilibrium quantum transport equation. The Berry
connection and Berry curvature is the fundamental physics \cite{fego} behind
the exact quantization of Hall conductance in units of $\frac{e^{2}}{h}$,
which also happens to coincide with the source and drain \textit{contact}
conductance per spin in a closed circuit of mesoscopic quantum transport
\cite{buot8}.

In Ref.\cite{previous}, we have shown that the $\left(  p.q;E.t\right)  $
phase-space is renormalized to that of $\left(  \mathcal{\vec{K}}%
,\mathcal{E}\right)  $ phase-space, where in the absence of magnetic fields,%
\begin{equation}
\mathcal{\vec{K}}=\vec{p}+e\vec{F}t, \label{eq1-1}%
\end{equation}
and
\begin{equation}
\mathcal{E}=E_{0}+e\vec{F}\cdot\vec{q} \label{eq2-2}%
\end{equation}
Here, the uniform electric field, $\vec{F}$, is in the $x$-direction, and the
Hall current is in the $y$-direction.

We have identified the topological invariant in $\left(  \mathcal{\vec{K}%
},\mathcal{E}\right)  $ phase-space nonequilibrium quantum transport equation
leading to IQHE. Moreover, the formula is also applicable to gapped
Landau-level structure of a free electron gas in intense magnetic field
\cite{kdp} since the variable $\mathcal{\vec{K}}$ can incorporates the
external vector potential, $\frac{e}{c}\vec{A}$, and its corresponding
parallel transport, if present. Note that the change of variables from
$\mathcal{\vec{K}}$ to $\vec{p}$ in the integration over the whole Brillouin
zone has a Jacobian unity.

It was shown in previous paper \cite{previous, covar} that the correct
expression for the IQHE conductivity given by%
\[
\ \sigma_{yx}=\frac{e^{2}}{h}\sum\limits_{\alpha}\frac{\Delta\phi_{total}%
}{2\pi}=\sum\limits_{\alpha}\frac{e^{2}}{h}n_{\alpha}%
\]

\subsection{Fractional quantum Hall effect}

Here, we give a simple explanation for FQHE effect conductance given by
\[
\sigma_{yx}=\frac{e^{2}}{h}v
\]
where $v$ is a fraction with prime number denominator in a hierarchy of
scaling factors. We remark that the number $2$ is a prime number and should
not be considered an even number. Moreover, prime integers exclude many odd
integers. All other non-prime integers can be decomposed into products of
prime numbers, this we refer to as\textit{ higher-order} scaling, i.e.,
higher-order power of primes. Thus, if one looks for a fundamental hierarchy
of scaling patterns, only \textit{fundamental} prime number scaling can
represent a primary hierarchy.

Although even denominators for $v$ are speculated in the literature, these do
not belong to a fundamental hierarchy of scaling in our present analysis for
the same reducibility or factorization reason into power of primes, and
therefore represent higher-order scaling. Moreover, although factorizable
scaling factors can appear in the measurements, these do not belong to a
fundamental hierarchy and we believe that these '\textit{higher-order}' type
of scaling hierarchy should appear very weak or less resolved in the
measurements. This is discussed in more detail in Sec.\ref{prime}.

\section{EFFECTS OF MAGNETIC FIELDS}

For free electrons, the presence of increasing magnetic field has two effects;
firstly it increases the energy levels in magnetic fields, $E_{n}=\hbar
\omega_{c}(n+1/2)$, $n=1,2,3,....$, where $\omega_{c}$ equals the cyclotron
frequency, $\frac{eB}{m}$, and their separations and secondly, it increases
the degeneracy of Landau levels and corresponding density of states, and hence
the flattening of the magnetic sub-band \cite{berg, wilk} of the lowest Landau level.

In the present paper, we show that the idea of flux attachment is inherent in
fully occupied as well as in partially occupied Landau levels, which results
in scaling factor closely related to the scaling $k$-factor in Chern-Simons
gauge theory. This is shown to be crucial in giving a simple explanation of
fractional quantum Hall effect (FQHE), using nonequilibrium quantum transport
in the lattice Weyl transform formalism \cite{bj,buot8} used in previous
papers \cite{previous}.

\subsection{Magnetic sub-bands}

\subsubsection{Low finite fields}

The sharp highly degenerate Landau levels for free electrons in a magnetic
field is broadened into magnetic energy sub-bands by the presence of periodic
atomic lattice sites. At values of the magnetic fields where the effect of the
periodic atomic lattice sites dominates, the effect of the magnetic fields can
be described simply in terms of the dynamics of the Bloch energy bands
\cite{wannier,zener} . For example, for finite magnetic fields, this is
manifested in de Haas-van Alphen effects \cite{devan,pekol} due to Landau
orbits at the Fermi surface, and magnetic breakdown between orbits in
complicated Fermi surfaces in metals \cite{bls,alex}. For smaller magnetic
fields, the Bloch-band dynamics is manifested in such phenomena as
paramagnetic and diamagnetic susceptibility. Indeed, interband coupling in
narrow-gap alloyed semimetals and semiconductors is responsible for the giant
diamagnetism of graphite, bismuth and Bi-Sb alloys \cite{bm, for}.

\subsubsection{Large magnetic fields}

However, when the effect of large magnetic fields dominates over the effect of
periodic atomic lattice sites, then the dynamics is now dictated by the
magnetic sub-bands. New length scales and time scales now dominate the
dynamics. These are the following,

\begin{center}%
\begin{tabular}
[c]{|r|r|}\hline
Cyclotron frequency & $\omega_{c}=\frac{eB}{m^{\ast}c}$\\\hline
Magnetic length & $l_{B}=\sqrt{\frac{\hbar c}{eB}}$\\\hline
Quantum flux & $\phi_{o}=\frac{2\pi\hbar c}{e}$\\\hline
Hall conductivity & $\sigma_{yx}=\frac{e^{2}}{2\pi\hbar}\nu$, \ \ ( $\nu$ is
an integer or a fraction)\\\hline
\end{tabular}

\end{center}

The major quantum mechanical operator in phase space is now played by
\[
\vec{K}=\vec{p}+\frac{e}{c}\vec{A}%
\]
which obeys the commutation relation%
\begin{equation}
\left[  K_{x},K_{y}\right]  =\frac{e\hbar}{ic}B_{z} \label{kapcommute}%
\end{equation}
This is more transparent if written as
\[
\left[  K_{x},K_{y}\right]  =\frac{e\hbar}{ic}B_{z}=\frac{e\hbar^{2}}{ic\hbar
}=-i\frac{\hbar^{2}}{l_{B}^{2}}%
\]
indicating that $l_{B}$ characterizes the wavelengths in a magnetic sub-bands.
For most magnetic field strengths, $l_{B}>a$ where $a$ is the atomic sites
lattice constant. This clearly supports the idea of magnetic sub-bands.

\subsubsection{von Neumann lattice in phase space}

Equation (\ref{kapcommute}) suggest the following relation%
\begin{equation}
\Delta K_{x}\Delta K_{y}\succeq\frac{1}{2}\frac{\hbar^{2}}{l_{B}^{2}}=\frac
{1}{2}\frac{e\hbar}{c}B_{z} \label{latticeA}%
\end{equation}
as a consequence of Heisenberg's uncertainty principle. Thus for the minimal
value of $\Delta K_{x}\Delta K_{y}=\frac{1}{2}\frac{\hbar^{2}}{l_{B}^{2}}$ a
von Neumann lattice in $2$-D phase space of $\left(  K_{x},K_{y}\right)  $ has
a lattice constant of the order $l_{B}$. For our purpose in what follows, it
is important to note that $\Delta K_{x}\Delta K_{y}\simeq\delta K_{x}\delta
K_{y}$ is proportional to the perpendicular magnetic field, $B_{z}$.
Therefore, we may also write a Poisson differential operator in a form,%
\begin{equation}
\frac{\partial^{\left(  a\right)  }}{\partial K_{x}}\frac{\partial^{\left(
b\right)  }}{\partial K_{y}}\sim\frac{1}{B_{z}}\text{,} \label{poisson}%
\end{equation}
as inversely proportional to the magnetic field. This equation means that if
we increase the magnetic field by a factor $k$, then from Eq. (\ref{poisson}),
we can scale the Poisson partial derivatives by $\frac{1}{k}$, resulting in
$\frac{1}{k}\left(  \frac{\partial^{\left(  a\right)  }}{\partial K_{x}}%
\frac{\partial^{\left(  b\right)  }}{\partial K_{y}}\right)  =\left(
\frac{\partial^{\left(  a\right)  }}{\partial K_{x}^{\prime}}\frac
{\partial^{\left(  b\right)  }}{\partial K_{y}^{\prime}}\right)  $ where the
primes embodied the updated magnetic field. This observation is crucial in our
macroscopic analysis of FQHE.

\section{LANDAU LEVEL DEGENERACY}

Classically, the Landau-level (LL) degeneracy may be approximated by%
\[
\frac{A}{\pi r^{2}}=\frac{R^{2}}{r^{2}}%
\]
where $r$ is the classical radius of tiny Landau orbits in a uniform magnetic
fields and the system area is given by $A=\pi R^{2}$. Equation (5) has the
dimensional units of total flux divided by the dimensional units of `quantum'
flux, $\frac{\left(  \pi R^{2}\right)  B}{\pi\left(  \frac{EL}{e}\right)
}\Rightarrow\sim\frac{\Phi}{\phi_{o}}$, where $E$ stands for units of energy.
In this classical analysis it is thus \textit{implied} that the magnetic flux
is equally divided or attached with each tiny cyclotron orbits of the same
energy for a fully occupied occupied Landau level.

Quantum mechanically, the more accurate expression for the Landau-level
degeneracy, $N_{LL}$ is $N_{LL}=\frac{\Phi}{\left(  \frac{2\pi\hbar c}%
{e}\right)  }=\frac{\Phi}{\phi_{o}}$. This can be inferred simply by a
Bohr-Sommerfeld quantization condition \cite{fego} which amounts to counting
of Planck states (`pixels' of action) in phase space using Berry's curvature
and connection, i.e., magnetic field and vector potential, respectively. In
Gaussian units, we have,%
\begin{align}
L\text{-}L\text{ }Degeneracy  &  =\frac{1}{2\pi\hbar}\iint\vec{\nabla}%
\times\vec{K}\cdot d\vec{a}\nonumber\\
&  =\frac{1}{2\pi\hbar}\iint\vec{\nabla}\times\frac{e}{c}\vec{A}\cdot d\vec
{a}\nonumber\\
&  =\frac{1}{2\pi\hbar}\frac{\left\vert e\right\vert }{c}\iint\vec{B}\cdot
d\vec{a}\nonumber\\
&  =\frac{\Phi}{\phi_{o}}\ \epsilon\
\mathbb{Z}
\text{, \ \ ({\small the same degeneracy for all Landau levels)}}\nonumber\\
&  =\frac{1}{2\pi\hbar}\oint\frac{e}{c}\vec{A}\cdot d\vec{q}=N_{LL}%
\ \epsilon\
\mathbb{Z}
, \label{b-s_version}%
\end{align}
where $\vec{K}=\vec{p}+$ $\frac{e}{c}\left(  \vec{A}+\vec{F}ct\right)  $
\cite{covar}, where $\vec{F}$ is the uniform electric field. Here, $\Phi$ is
the total magnetic flux, and $\phi_{o}=\frac{hc}{\left\vert e\right\vert }$ is
the quantum flux which one may considered as \textit{quantum mechanically}
\textit{attached} to each quasiparticle (note that instead of classical tiny
cyclotron orbits, we now refer these as \textit{quasiparticles}) in a fully
occupied lowest Landau level (LLL). This is actually a precursor of the
concept of \textit{flux attachment} afforded by the Chern-Simons $U\left(
1\right)  $ gauge theory in $2$-dimensions. Moreover, this is also realization
of the B-S quantization condition given in Eq. (\ref{b-s_version})
\cite{fego}. The resulting quantization of orbital motion leads to edge states
and integer quantum Hall effect under uniform magnetic fields. The general
analysis of edge states marks the works of Laughlin \cite{laughlin, laugh2},
and Halperin \cite{halperin}.

\subsubsection{Flux attachments}

From Eq. (\ref{b-s_version}), we have for a fully occupied lowest Landau level
at $v=1$ of the experiments,%
\begin{equation}
\frac{1}{2\pi\hbar}\iint\vec{\nabla}\times\vec{K}\cdot d\vec{a}=\frac{\Phi
}{\phi_{o}}=N_{o}=N_{LLL} \label{eqmotion}%
\end{equation}
where as before, $N_{o}$ is the total number of available electrons, e.g., for
a given gate voltage bias of a Si MOSFET or GaAs MESFET heterostructures, and
$N_{LLL}$ is the degeneracy of the LLL.

For fractionally occupied, say $\frac{1}{3}$ of LL at higher magnetic fields
or higher degeneracies, $\frac{N_{LLL}}{N_{LL}}=\frac{N_{0}}{N_{LL}}=\frac
{1}{3}$, so we can simply re-express Eq (\ref{eqmotion}), on the spirit of the
flux-attachment concept as%
\begin{equation}
\frac{k}{2\pi\hbar}\iint\vec{\nabla}\times\vec{K}\cdot d\vec{a}=\frac{\Phi
}{\phi_{o}}=kN_{o}=N_{LL} \label{kfactor}%
\end{equation}
where we have indicated the scaling factor by $k$, e.g., $k=3.$

\subsubsection{Either flux attachments or fractional charge}

Equation (\ref{kfactor}) means that naively, either $3\phi_{o}=\frac
{3hc}{\left\vert e\right\vert }$ is attach to each quasiparticle or that
$\left\vert e\right\vert $ in $\phi_{o}=\frac{hc}{\left\vert e\right\vert }$
is $\frac{\left\vert e\right\vert }{3}$, a fractional charge. These two
alternatives carry different physical meanings, flux attachment means the
number of electrons is not enough to fully occupy the new density of states or
LL degeneracy, whereas fractional charge means the new quasiparticles fully
occupy the new Landau level degeneracy, $N_{LL}$, but with fractional charge.

\subsection{Hierarchy of primes for the scaling factor $k$}

\label{prime}

Since the magnetic flux of the LLL is our reference point for scaling by a
factor $k$, in order to form a hierarchy of scaling, this cannot be expressed
as product of prime numbers, otherwise we will eventually be scaling a
different magnetic flux after the first factor of the product. Thus,
factorizable numbers and fractions cannot represent as members of a
fundamental hierarchy of scaling factor $k$. Thus, for integer $k$ this must
be a prime number, whereas for fractions the numerator and denominator of $k$
must also be a prime numbers. This scaling should constitute the principal
scaling hierarchy, i.e., first-order in primes.

A second scaling hierarchy would constitute prime numbers for integers and
prime-number numerators only for $k$. The third type of scaling hierarchy will
constitute prime number denominators only for $k$. Since all these scaling $k$
are factorizable, we consider these as '\textit{higher-order}' scaling
hierarchy and to be weakly observed. The fourth scaling hierarchy is for both
numerators and denominators to be factorizable. This we expect to be rarely
seen in the experiments.

\subsubsection{Higher probability for prime-number numerators for $k$}

Therefore, from the fundamental and second scaling hierarchies, we expect the
dominance of prime-number numerators for $k$ in the experiments. We summarize
these statements by the following table which give some examples of the values
of $k$ and its inverse,%

\begin{equation}%
\begin{tabular}
[c]{|r|r|}\hline
\multicolumn{2}{|r|}{$k\geq1$}\\\hline
$k$ & $\nu=\frac{1}{k}$\\\hline
$1$ & $\nu=1$\\\hline
$2$ & $\nu=\frac{1}{2}$\\\hline
$3$ & $\nu=\frac{1}{3}$\\\hline
$5$ & $\nu=\frac{1}{5}$\\\hline
$7$ & $\nu=\frac{1}{7}$\\\hline
$\frac{3}{2}$ & $\nu=\frac{2}{3}$\\\hline
$\frac{5}{3}$ & $\nu=\frac{3}{5}$\\\hline
$\frac{5}{4}$ & $\nu=\frac{4}{5}$\\\hline
$\frac{7}{2}$ & $\nu=\frac{2}{7}$\\\hline
$\frac{7}{3}$ & $\nu=\frac{3}{7}$\\\hline
$\frac{7}{4}$ & $\nu=\frac{4}{7}$\\\hline
$\frac{7}{5}$ & $\nu=\frac{5}{7}$\\\hline
$..$ & $...$\\\hline
\end{tabular}
\ \ \ \ \ \ \ \ \ ,\text{ \ }%
\begin{tabular}
[c]{|r|r|}\hline
\multicolumn{2}{|r|}{$k<1$}\\\hline
$k$ & $\nu=\frac{1}{k}$\\\hline
$\frac{2}{3}$ & $\nu=\frac{3}{2}$\\\hline
$\frac{2}{5}$ & $\nu=\frac{5}{2}$\\\hline
$\frac{2}{7}$ & $\nu=\frac{7}{2}$\\\hline
$\frac{3}{4}$ & $\nu=\frac{4}{3}$\\\hline
$\frac{3}{5}$ & $\nu=\frac{5}{3}$\\\hline
$\frac{3}{6}$ & $\nu=2$\\\hline
$\frac{3}{7}$ & $\nu=\frac{7}{3}$\\\hline
$\frac{5}{6}$ & $\nu=\frac{6}{5}$\\\hline
$\frac{5}{7}$ & $\nu=\frac{7}{5}$\\\hline
$..$ & $..$\\\hline
$..$ & $..$\\\hline
$..$ & $..$\\\hline
$..$ & $...$\\\hline
\end{tabular}
\ \ \ \ \ \ \ \ \ \label{kvalues}%
\end{equation}
We remark that the number $2$ in $\nu=\frac{1}{2}$ is a prime number and
should not be considered an even number. The whole number in $\nu=2$ in Eq.
(\ref{kvalues}) simply means two LL are filled by the reference population of
the LLL, $N_{o}=N_{LLL}$, and corresponds to the re-imergence of IQHE.
Moreover, prime integers exclude many odd integers. Thus, it appears that the
assumption in the literature that the denominator of $v$ is given by the
expression, $\left(  2n+1\right)  $, in a hierarchy is wrong, this claim is
simply borne out of the dominance of prime number numerators for $k$ ($2$ is a
prime number). Furthermore, although even denominators for $v$ are speculated
in the literature, these cannot be members of a fundamental scaling hierarchy
for the same reason that these can be decomposed into products of prime
numbers. The entries of Eq. (\ref{kvalues}) actually exist as experimental
values \cite{stormer1,stormer2,stormer3}, see Figs. \ref{fig2} and \ref{fig3}.

\section{RELATION TO CHERN-SIMONS GAUGE THEORIES}

We write the Chern-Simons Lagrangian density for $U\left(  1\right)  $ gauge
theory for $2$-dimensional system of manifold, $M$, as%
\begin{equation}
\mathcal{L}_{CS}=\gamma\varepsilon^{\mu\lambda\nu}A_{\mu}\partial_{\lambda
}A_{\nu}-A_{\mu}J^{\mu} \label{Lcs}%
\end{equation}
where $J^{\mu}=\left(  \rho,\vec{J}\right)  $, $\rho$ is the charge density
and $\vec{J}$ is the current density. Later, we will associate the parameter
$\gamma$ with our scaling parameter. Equation (\ref{Lcs}) is often referred to
as the Maxwell Chern-Simons theory. The equation of motion is obtained by
variation with respect to $A_{\mu}$%
\[
\frac{\delta\mathcal{L}_{CS}}{\delta A_{\mu}}=\gamma\varepsilon^{\lambda\nu
}\partial_{\lambda}A_{\nu}-J^{\mu}=0
\]
This gives%
\begin{equation}
\gamma%
{\displaystyle\int\limits_{M}}
\nabla\times\vec{A}=%
{\displaystyle\int\limits_{M}}
J^{0}=%
{\displaystyle\int\limits_{M}}
\rho\label{eqmotion2}%
\end{equation}
For a fully occupied LLL, we may equate the following
\begin{align*}
\gamma &  =\frac{1}{2\pi\hbar}\frac{e}{c}\\
\nabla\times\vec{A}  &  \Longrightarrow\nabla\times\frac{e}{c}\vec{A}%
=\nabla\times\vec{K}.\\%
{\displaystyle\int\limits_{M}}
\rho &  =N_{LL}%
\end{align*}
where $\vec{K}=P+$ $\frac{e}{c}\vec{A}+\vec{F}ct$, \cite{covar} where $\vec
{F}$ is the uniform electric field. Thus, Eq. (\ref{eqmotion2}) reduces to
that of Eq. (\ref{eqmotion}). The relation of Eq. (\ref{kfactor}) with
Chern-Simons $U\left(  1\right)  $ gauge theory thus becomes clear. We
re-write Eq. (\ref{kfactor}) with $k=3$ as%
\begin{equation}
\frac{k}{2\pi\hbar}\iint\vec{\nabla}\times\frac{e}{c}\vec{A}\cdot d\vec
{a}=\frac{\Phi}{\phi_{o}}=N_{LL} \label{kfactor2}%
\end{equation}
where $\vec{A}$ is the vector potential. Equation (\ref{kfactor2}) clearly
shows that the magnetic field is scaled by an integer factor $k$ to yield a
larger number, $N_{LL}$, of degenerate states in a Landau level compared to
our reference fully occupied LLL, resulting in fractional filling. Equation
(\ref{Lcs}), with $k\gamma=k\left(  \frac{e}{2\pi\hbar c}\right)  $, is
basically in the Chern-Simons form of $U\left(  1\right)  $ gauge theory.

\subsection{Experiments in IQHE and FQHE}

\begin{figure}
[h]\centering
\includegraphics[width=2.0167in]{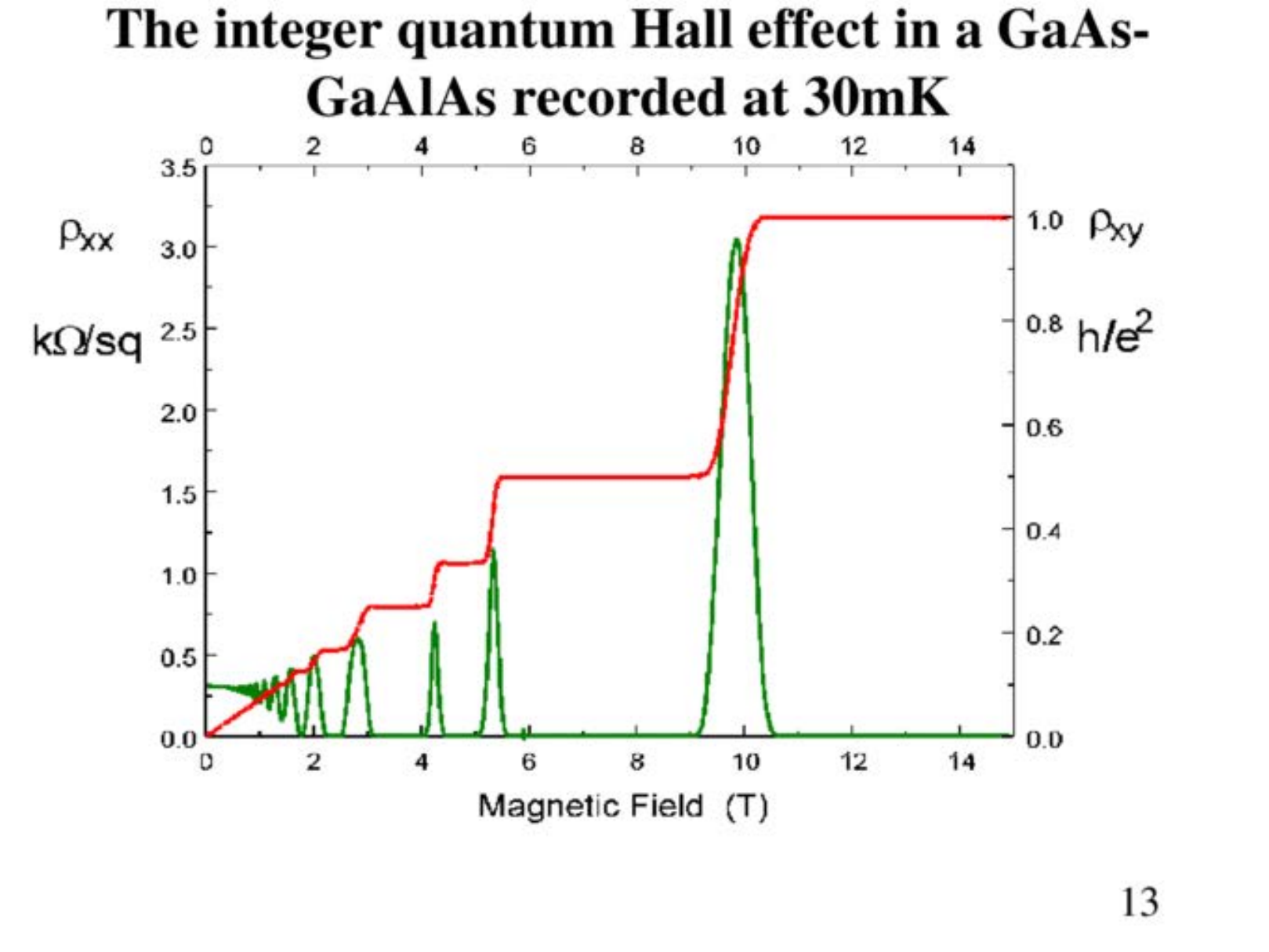}
\caption{The experimental IQHE in GaAs/AlGaAs first observed by von Klitzing,
Dorda and Pepper \cite{kdp}.} \label{fig1}
\end{figure}

In 1980, Klaus von Klitzing, working with Si MOSFET samples developed by
Michael Pepper and Gerhard Dorda, made the unexpected discovery, see Fig.
\ref{fig1}, that the Hall resistance was exactly quantized \cite{kdp}. Other
experiments are done in GaAs MESFETs.

\begin{figure}
[h]\centering
\includegraphics[width=2.77in]{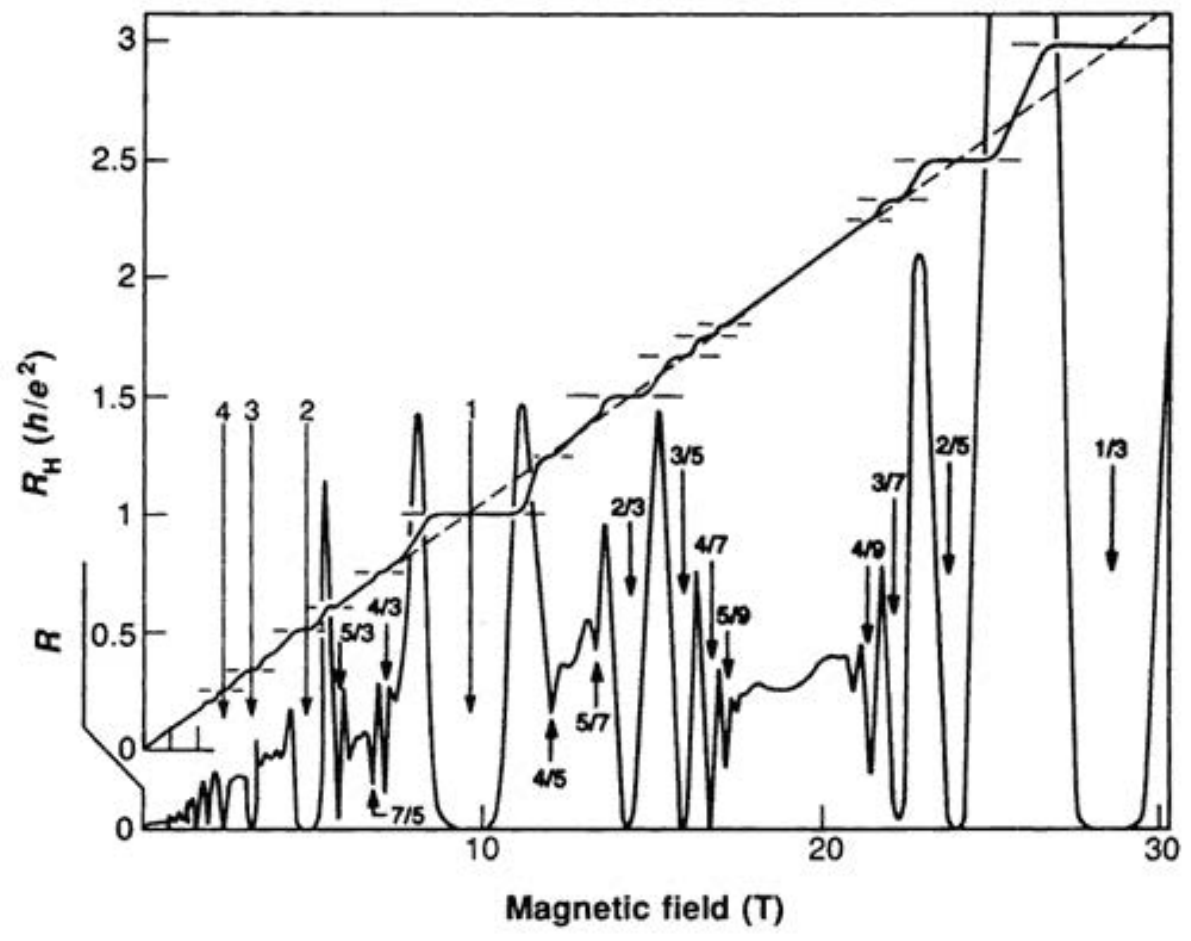}
\caption{The dashed diagonal line represents the classical Hall resistance and
the full drawn diagonal stepped curve the experimental results. The magnetic
fields causing the steps are marked with arrows. The step first discovered by
St\"{o}rmer and Tsui \cite{stormer1} at the highest value of the magnetic
field and the steps earlier discovered by von Klitzing (integers) with a
weaker magnetic field.. Here we consider the $v=1$ as our reference in
calculating the $v=\frac{1}{3}$ by employing our novel approach.} \label{fig2}
\end{figure}

In 1982, a group of physicists lead by Tsui and Stormer, working on GaAs
MESFET heterostructures, discovered $\nu=\frac{1}{3}$ state \cite{stormer1}.
This means they found a state in which the FQHE conductivity was given by%
\[
\sigma_{yx}=\frac{e^{2}}{2\pi\hbar}v
\]
as shown in Fig. \ref{fig2}. In general, it was discovered over time that
there were several states with fractional values that had plateaus
corresponding to values of quantum Hall resistivity that correspond to
different $\nu$'s. This was confusing how these states could occur since
$\nu=1$ should be the lowest possible state given the scheme found by the
Landau levels above. At that time, it was considered strange since some
fractions would occur and others would not. For example there was $\nu
=\frac{1}{3}$ but not $\nu=\frac{1}{2}$ . There's $\nu=\frac{3}{5}$ , but
there is no $\nu=\frac{3}{4}$ . In general, it was found at first that no
state had an even denominator.

\begin{figure}
[h]\centering
\includegraphics[width=2.8885in]{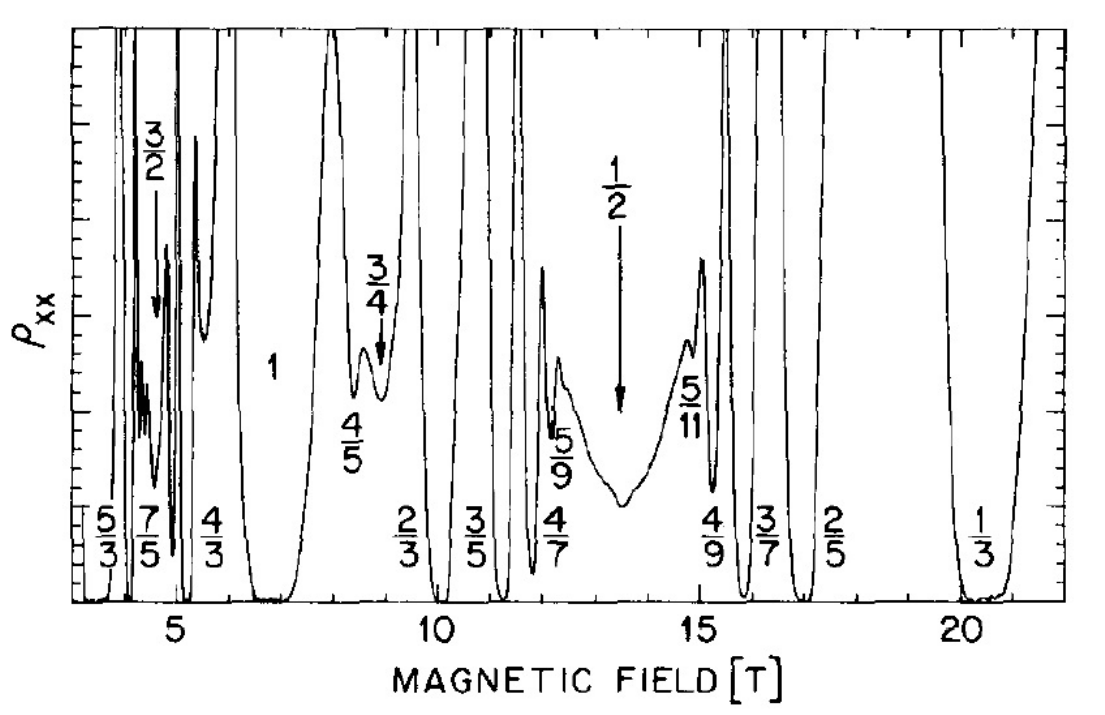}
\caption{Graph showing the diagonal resistivity. $\rho_{xx}$, at T = 80 mK
showing strong transport anomalies at $\nu=\frac{1}{2}$ and $\nu=\frac{3}{2}$
[Reproduced from Ref. \cite{stormer2}].} \label{fig3}
\end{figure}

\begin{figure}
[h]\centering
\includegraphics[width=4.101800in]{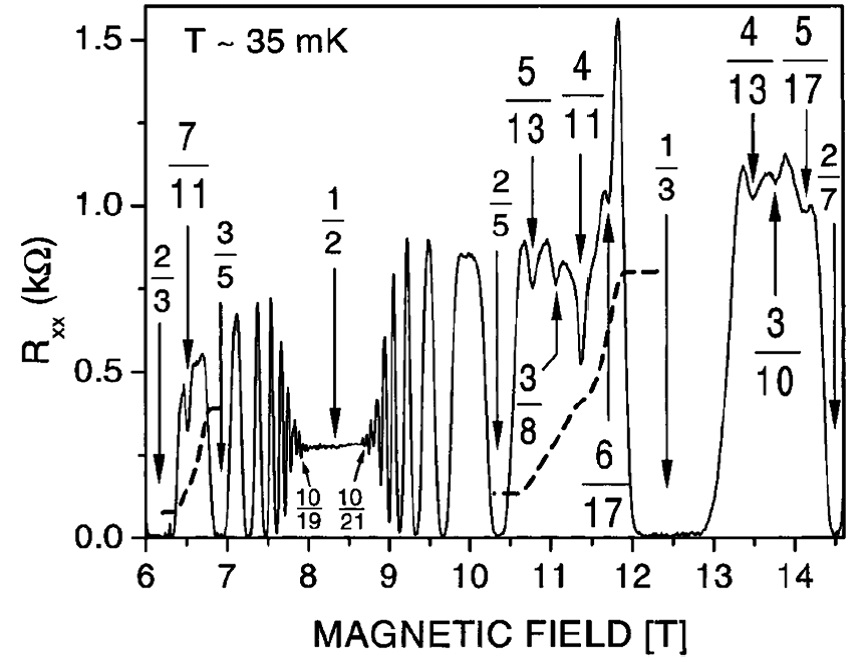}
\caption{In the present analysis, $v=\frac{10}{21},\frac{3}{8},$and $\frac
{3}{10}$, which are weakly resolved in the figure, do not belong to a
hierarchy of scaling factors, since the denominators are factorizable and not
prime numbers. Figure reproduced from Ref.\cite{stormer3}} \label{fig4}
\end{figure}

\begin{figure}
[h]\centering
\includegraphics[width=3.967800in]{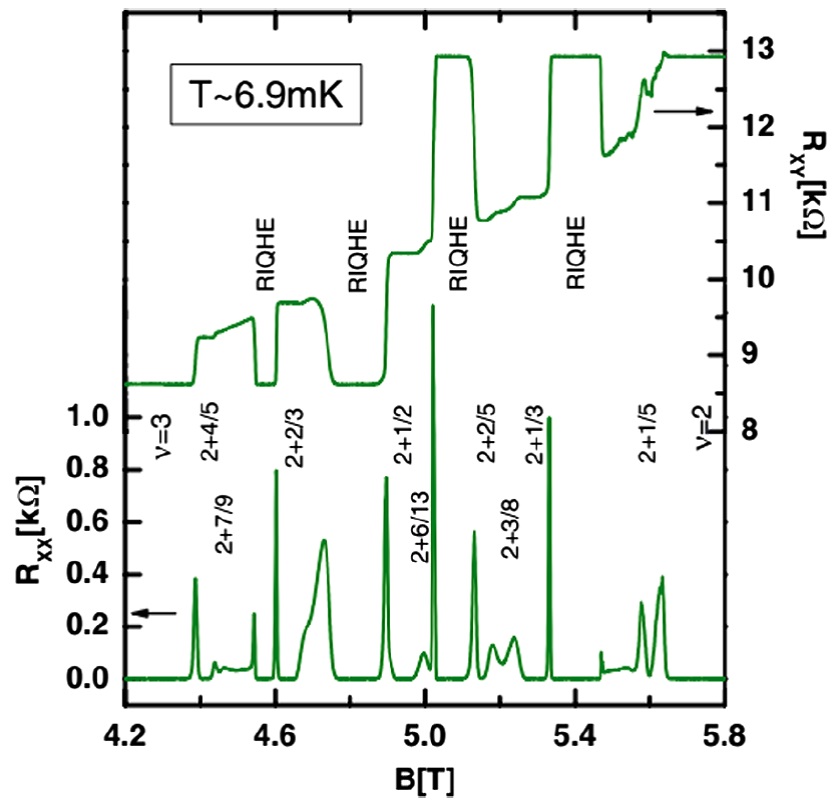}
\caption{A quantum Hall data showing plateaus in the second Landau level. Note
that many of the plateaus are labeled RIQHE (reentrant integer quantum Hall
effect). Note that only prime number denominators are resolved. Figure
reproduced From Ref.\cite{kumar}} \label{fig5}
\end{figure}

However, more later experiments have presented evidence for even denominator,
see Fig. \ref{fig3}, in some expecial cases \cite{stormer2}. More detailed
experiments were done \cite{stormer3, kumar}, see Figs. \ref{fig4} and
\ref{fig5}.

Our analysis in this paper is not based on variational wavefunction
microscopic approaches, for example, using Laughlin wavefunctions
\cite{laugh2}, More-Read wavefunctions \cite{more}, or Jain wavefunctions
\cite{jain, jain2, jain3}, but is based on macroscopic phase-space analysis
reminiscent of the macroscopic Landau-Ginsburg analysis of phase transitions
in matter via symmetry breaking, employing minimal description using order
parameters. Here, our '\textit{order}' parameter is the magnetic flux,
$\frac{\Phi}{\phi_{o}}$, or more precisely the prime number \textit{scaling
factor} of the magnetic flux of a fully occupied lowest Landau level (LLL) for
a given geometrical $2$-D feature size, i.e., channel area, of either Si
MOSFET or GaAs MESFET heterostructures used in the experiments
\cite{stormer1,stormer2, stormer3}.

Further experiments in GaAs/AlGaAs MESFET revealed strong dominance of prime
number denominators for the filling factor $\nu$, see Figs. \ref{fig4} and
\ref{fig5}$.$

Another interesting approach to FQHE, which makes use of Bohr-Sommerfeld
quantization, was given by Jacak \cite{jacak1,jacak2}. This approach may be
related to our use here of Maxwell Chern-Simons $U\left(  1\right)  $ gauge
theory, which may also related to the Bohr-Sommerfeld quantization \cite{fego}.

\section{TOPOLOGICAL INVARIANT IN PHASE SPACE AND FQHE}

The following transport analysis give a simple account of the experimental
results \cite{stormer1, stormer2,stormer3} with the fully occupied lowest
Landau level as a \textit{reference point}, i.e., the point $v=1$ moving
towards successive $v$'s$\,<1$ with increasing magnetic fields, or moving
backward of successive $v$'s$\,>1$ with decreasing magnetic fields. In
particular, we will focus on $v=\frac{1}{3}$ \ or $k=3$ as a particular case
since this looks like a very defined state in the experiments, although the
following analysis holds for any values of $k$.

The topologically invariant result for $\sigma_{yx}$ will thus come out to be,%
\begin{align}
&  e^{2}\left\vert \vec{F}\right\vert \frac{1}{\left(  2\pi\hbar\right)  ^{2}}%
{\displaystyle\int}
{\displaystyle\int}
{\displaystyle\int}
dk_{x}dk_{y}dt\ \left(  \frac{1}{3}\right)  \left[  \frac{\partial^{\left(
a\right)  }}{\partial\mathcal{\vec{K}}_{x}}\frac{\partial^{\left(  b\right)
}}{\partial\mathcal{\vec{K}}_{y}}-\frac{\partial^{\left(  a\right)  }%
}{\partial\mathcal{\vec{K}}_{y}}\frac{\partial^{\left(  b\right)  }}%
{\partial\mathcal{\vec{K}}_{x}}\right] \nonumber\\
&  \times H^{\left(  a\right)  }\left(  \mathcal{\vec{K}}^{\prime}%
,\mathcal{E}\right)  \left(  -iG^{<\left(  b\right)  }\left(  \mathcal{\vec
{K}}^{\prime},\mathcal{E}\right)  \right)  \label{scaledtopo}%
\end{align}
where $\mathcal{\vec{K}}^{\prime}$ comes from the actual lattice Weyl
tranformation with updated magnetic field. The factor $\left(  \frac{1}%
{3}\right)  $ comes from the relation of Eq. (\ref{poisson}), with $k=3$. Note
that the integrals have to do with the counting of the number of occupied
states, moderated by the energy-dependent Wigner distribution, $-iG^{<\left(
b\right)  }\left(  \mathcal{\vec{K}}^{\prime},\mathcal{E}\right)  $, of
particle population in phase space.

Thus, in place of the previous expression for fully occupied LLL for $k=1$, we
now have, for any value of $k,$
\begin{align}
\sigma_{yx}  &  =\left(  \frac{1}{k}\right)  \frac{e^{2}}{\left(  2\pi
\hbar\right)  ^{2}}%
{\displaystyle\int}
{\displaystyle\int}
{\displaystyle\int}
dp_{x}dp_{y}dt\ \left[  \frac{\partial^{\left(  a\right)  }}{\partial
\mathcal{\vec{K}}_{x}}\frac{\partial^{\left(  b\right)  }}{\partial
\mathcal{\vec{K}}_{y}}-\frac{\partial^{\left(  a\right)  }}{\partial
\mathcal{\vec{K}}_{y}}\frac{\partial^{\left(  b\right)  }}{\partial
\mathcal{\vec{K}}_{x}}\right] \nonumber\\
&  \times H^{\left(  a\right)  }\left(  \mathcal{\vec{K}}^{\prime}%
,\mathcal{E}\right)  \left(  -iG^{<\left(  b\right)  }\left(  \mathcal{\vec
{K}}^{\prime},\mathcal{E}\right)  \right)  \label{hallcond}%
\end{align}

\subsubsection{Magnetic sub-band representation of magnetic quantum states}

Now the common representation of $\mathcal{\vec{K}}$ and $\mathcal{\vec{K}%
}^{\prime}$ of the differential operator in Eq. (\ref{scaledtopo}) is the
$\vec{p}$ of their magnetic sub-bands. Therefore this will yield the same
result if we write as
\[
\left[  \frac{\partial^{\left(  a\right)  }}{\partial p_{x}}\frac
{\partial^{\left(  b\right)  }}{\partial p_{y}}-\frac{\partial^{\left(
a\right)  }}{\partial p_{y}}\frac{\partial^{\left(  b\right)  }}{\partial
p_{x}}\right]  H^{\left(  a\right)  }\left(  \mathcal{\vec{K}}^{\prime
},\mathcal{E}\right)  \left(  -iG^{<\left(  b\right)  }\left(  \mathcal{\vec
{K}}^{\prime},\mathcal{E}\right)  \right)  ,
\]
since%
\[
\frac{\partial^{\left(  a\right)  }}{\partial\mathcal{\vec{K}}_{x}}\equiv
\frac{\partial\mathcal{\vec{K}}_{x}}{\partial p_{x}}\frac{\partial^{\left(
a\right)  }}{\partial\mathcal{\vec{K}}_{x}}=\frac{\partial^{\left(  a\right)
}}{\partial p_{x}}%
\]
and so on.

Therefore, by going over the analogous procedure used before \cite{previous},
following Eq. (\ref{scaledtopo}) we finally obtain,%
\begin{equation}
\sigma_{yx}=\left(  \frac{e^{2}}{h}\right)  v\text{ \ \ \ }\left(  v=\frac
{1}{3}\right)  \label{FQHE}%
\end{equation}
Generalizing, we have%
\begin{equation}
\sigma_{yx}=\left(  \frac{e^{2}}{h}\right)  v\text{ },\text{ \ \ \ }\left(
v=\frac{1}{k}\right)  \label{FQHE2}%
\end{equation}

For convenience, the derivation is given in some details in the Appendix A.
The readers is also referred to Ref. \cite{previous} for more details. In Eq.
(\ref{FQHE}), there is no summation over occupied levels since we are here
mainly concerned with scaling hierarchy of fractionally occupied lowest Landau
level with increasing magnetic field beyond the field at the fully occupied LLL.

\section{CONCLUDING REMARKS}

Our analysis of FQHE parallels that of our previous papers \cite{previous,
comments, jagna, covar} on IQHE . The only new theoretical ingredient that we
need is Eq. (\ref{b-s_version}) for the magnetic field, which is equivalent to
the Chern-Simons $U\left(  1\right)  $ gauge theory equation of motion, Eq.
(\ref{kfactor2}). The $k$-parameter essentially serves as our order parameter
analogous to Landau-Ginsburg order parameter. This naturally brings in the
flux attachment concept that is concomitant to new statistics of
quasiparticles, the so-called anyons. Physically, this means that instead of
talking about filled gapped energy levels in IQHE, we are treating
fractionally filled gapped energy levels at different values of the magnetic
field, specifically using a fully filled LLL as the point of reference in FQHE
analysis of this paper. \textit{This will naturally produced hierarchies of
scaling factors, }$k$. We observe that experiments are done on
heterostructured semiconductors, e.g., Si MOSFET or GaAs MESFET, where
electrons can be considered $2$-D Bloch electrons in magnetic sub-bands.

We have shown that the scaling $k$-factor in Chern-Simons gauge theory
corresponds to the scaling of the magnetic fields in FQHE. This leads us to
account for the flattening or deformation of the magnetic sub-band of the LLL
upon increasing the magnetic field. Following similar procedure using the
nonequilibirum quantum transport in a lattice Weyl transform formalism, we
determined the exact expression for the Hall conductivity in FQHE found in experiments.

Since the magnetic flux of the LLL is our reference point for scaling by a
factor $k$, this cannot be expressed as product of numbers within the
fundamental hierarchy, otherwise we will eventually be scaling a different
magnetic flux after the first factor of the product. Thus, the integer $k$
this must be a prime number, whereas for fractions the numerator and
denominator of $k$ must also be prime numbers. Indeed, the experiments clearly
show that the value of the magnetic flux scaling number $k$ has a prime number
numerator, or FQHE scaling fraction $\nu$ with prime number denominator. These
are what is expected to have higher probability of occurring and is mostly
seen in experiments. Prime integers exclude many odd integers. Thus, it
appears that the assumption in the literature that the denominator of $v$ is
given by the expression, $\left(  2n+1\right)  $, in a fundamental scaling
hierarchy is wrong. Furthermore, although even denominators for $v$ are
speculated in the literature, these cannot be members of a fundamental
hierarchy for reason of factorizability. Therefore, these prime numbers in our
theory must be the controlling parameters of stable FQHE conductance
well-resolved in the experiments \cite{kumar}.

\begin{acknowledgement}
One of the authors (FAB) is grateful for 'Balik Scientist' Program Award of
PCIEERD-DOST, Philippines, for the Visiting Professorship at the CTCMP, Cebu
Normal University.
\end{acknowledgement}

\appendix

\section{REVERTING TO MATRIX ELEMENTS}

\subsection{Lattice Weyl transform}

We refer the readers to \cite{previous, comments, jagna, covar} for more
details of this section. Denoting the operation of taking the lattice Weyl
transform by the symbol $\mathcal{W}$ then the lattice Weyl transform of total
partial derivatives is given by \cite{previous},
\begin{align}
\mathcal{W}\left(  \frac{\partial}{\partial q^{\prime}}+\frac{\partial
}{\partial q^{\prime\prime}}\right)  \left\langle q^{\prime},\lambda^{\prime
}\right\vert \mathbf{\hat{A}}\left\vert q^{\prime\prime},\lambda
^{^{\prime\prime}}\right\rangle  &  =\frac{\partial}{\partial q}%
\sum\limits_{v}e^{\left(  \frac{2i}{\hbar}\right)  p\cdot v}\left\langle
q-v,\lambda^{\prime}\right\vert \mathbf{\hat{A}}\left\vert q+v,\lambda
^{\prime\prime}\right\rangle ,\nonumber\\
&  =\frac{\partial}{\partial q}A_{\lambda^{\prime}\lambda^{\prime\prime}%
}\left(  p.q\right)  \text{.} \label{sumq}%
\end{align}
Similarly
\begin{align}
\mathcal{W}\left(  \frac{\partial}{\partial p^{\prime}}+\frac{\partial
}{\partial p^{\prime\prime}}\right)  \left\langle p^{\prime},\lambda^{\prime
}\right\vert \mathbf{\hat{A}}\left\vert p^{\prime\prime},\lambda
^{^{\prime\prime}}\right\rangle  &  =\frac{\partial}{\partial p}%
\sum\limits_{u}e^{\left(  \frac{2i}{\hbar}\right)  q\cdot u}\left\langle
p+u,\lambda\right\vert \mathbf{\hat{A}}\left\vert p-u,\lambda^{\prime
}\right\rangle ,\nonumber\\
&  =\frac{\partial}{\partial p}A_{\lambda^{\prime}\lambda^{\prime\prime}%
}\left(  p.q\right)  \text{.} \label{sump}%
\end{align}
In applying to problems in uniform electromagnetic fields, the form
\begin{equation}
\left\langle \vec{q}_{1},t_{1}\right\vert \hat{H}^{\left(  1\right)
}\left\vert \vec{q}_{2},t_{2}\right\rangle \Longrightarrow e^{-i\frac{e}{\hbar
c}\left(  \vec{A}\left(  q\right)  +\vec{F}ct\right)  \cdot\left(  \vec{q}%
_{1}-\vec{q}_{2}\right)  }e^{-i\frac{e}{\hbar}\vec{F}\cdot\vec{q}\left(
t_{1}-t_{2}\right)  }H^{\left(  1\right)  }\left(  \vec{q}_{1}-\vec{q}%
_{2},t_{1}-t_{2}\right)  \text{,} \label{peierlsPhase}%
\end{equation}
where%
\begin{align*}
\vec{q}  &  =\frac{1}{2}\left(  \vec{q}_{1}+\vec{q}_{2}\right)  \text{,}\\
t  &  =\frac{1}{2}\left(  t_{1}+t_{2}\right)  \text{.}%
\end{align*}
of matrix elements in Eq. (\ref{peierlsPhase}), we have%
\begin{align}
&  \left\langle \vec{q}-\vec{v};t-\frac{\tau}{2},\lambda\right\vert
\mathbf{\hat{A}}\left\vert \vec{q}+\vec{v};t+\frac{\tau}{2},\lambda^{\prime
}\right\rangle \nonumber\\
&  =e^{-i\frac{e}{\hbar}\vec{F}t\cdot\left(  \vec{q}_{1}-\vec{q}_{2}\right)
}e^{-i\frac{e}{\hbar}\vec{F}\cdot\vec{q}\left(  t_{1}-t_{2}\right)  }A\left(
\vec{q}_{1}-\vec{q}_{2},t_{1}-t_{2}\right) \nonumber\\
&  =e^{i\frac{e}{\hbar}\vec{F}t\cdot\left(  2\vec{v}\right)  }e^{i\frac
{e}{\hbar}\vec{F}\cdot\vec{q}\tau}A_{\lambda\lambda^{\prime}}\left(  \vec
{q}_{1}-\vec{q}_{2},t_{1}-t_{2}\right)  \text{.} \label{matrix}%
\end{align}
Thus%
\begin{align}
A_{\lambda\lambda^{\prime}}\left(  \vec{p}.\vec{q};E,t\right)   &
=\sum\limits_{\vec{v};\tau}e^{\left(  \frac{2i}{\hbar}\right)  \left(  \vec
{p}+\frac{e}{x}\vec{A}\left(  \vec{q}\right)  \right)  \cdot2\vec{v}%
}e^{\left(  \frac{i}{\hbar}\right)  E\tau}e^{i\frac{e}{\hbar}\vec{F}%
t\cdot\left(  2\vec{v}\right)  }e^{i\frac{e}{\hbar}\vec{F}\cdot\vec{q}\tau
}A_{\lambda\lambda^{\prime}}\left(  \vec{q}_{1}-\vec{q}_{2},t_{1}%
-t_{2}\right)  ,\nonumber\\
&  =\sum\limits_{\vec{v};\tau}e^{\left(  \frac{2i}{\hbar}\right)  \left(
\vec{p}+e\vec{F}t\right)  \cdot\vec{v}}e^{\left(  \frac{i}{\hbar}\right)
\left(  E+e\vec{F}\cdot\vec{q}\right)  \tau}A_{\lambda\lambda^{\prime}}\left(
\vec{q}_{1}-\vec{q}_{2},t_{1}-t_{2}\right)  ,\nonumber\\
&  =A_{\lambda\lambda^{\prime}}\left(  \left(  \vec{p}+\frac{e}{c}\left(
\vec{A}+\vec{F}ct\right)  \right)  ;\left(  E+e\vec{F}\cdot q\right)  \right)
,\nonumber\\
&  =A_{\lambda\lambda^{\prime}}\left(  \mathcal{\vec{K}};\mathcal{E}\right)
\text{.} \label{matrixphase}%
\end{align}
Hence the relevant dynamical variables in the phase space including the time
variable occurs in particular combinations of $\mathcal{\vec{K}}$ and
$\mathcal{E}$. Therefore, besides \ the crystal momentum varying in time as
\begin{equation}
\mathcal{\vec{K}}=\vec{p}+\frac{e}{c}\left(  \vec{A}+\vec{F}ct\right)
\text{,} \label{shiftedK}%
\end{equation}
the energy variable vary with $\vec{q}$ due to the electric field as
\begin{equation}
\mathcal{E}=E_{o}+e\vec{F}\cdot\vec{q}. \label{shiftedE}%
\end{equation}

Note that the derivatives on the LHS of Eqs. (\ref{sumq}) and (\ref{sump})
obviously operate only on the wavefunctions or state vectors not on the
operator. In Eqs. (\ref{sump}) and (\ref{sump}), we use the four dimensional
notation, e.g, $q=\left(  \vec{q},t\right)  $, and $p=\left(  \vec
{p},E\right)  .$

\subsubsection{Derivation of FQHE}

The Hall current in the $y$-direction maybe written as \cite{previous,
comments, jagna, covar}, after dividing by $\frac{1}{k}$ the Poisson operator
as,
\begin{align}
&  \frac{a^{2}}{\left(  2\pi\hbar\right)  ^{2}}\int\int dp_{x}dp_{y}\left(
\frac{e}{a^{2}}\frac{\partial\mathcal{E}}{\partial\mathcal{\vec{K}}%
_{y}^{\prime}}\right)  \left(  -iG^{<}\left(  \mathcal{\vec{K}}^{\prime
},\mathcal{E}\right)  \right) \nonumber\\
&  =e^{2}\left\vert \vec{F}\right\vert \frac{1}{\left(  2\pi\hbar\right)
^{2}}\int\int\int dp_{x}dp_{y}dt\nonumber\\
&  \times\ \left(  \frac{1}{k}\right)  \left[  \frac{\partial^{\left(
a\right)  }}{\partial\mathcal{\vec{K}}_{x}}\frac{\partial^{\left(  b\right)
}}{\partial\mathcal{\vec{K}}_{y}}-\frac{\partial^{\left(  a\right)  }%
}{\partial\mathcal{\vec{K}}_{y}}\frac{\partial^{\left(  b\right)  }}%
{\partial\mathcal{\vec{K}}_{x}}\right]  H^{\left(  a\right)  }\left(
\mathcal{\vec{K}}^{\prime},\mathcal{E}\right)  \left(  -iG^{<\left(  b\right)
}\left(  \mathcal{\vec{K}}^{\prime},\mathcal{E}\right)  \right)  \text{.}
\label{eqKE}%
\end{align}

Now to show that the Eq. (\ref{hallcond}) gives $\sigma_{yx}=\frac{e^{2}}%
{h}\nu$, where $\nu=\frac{1}{k}$, we need to transform the integral of Eq.
(\ref{hallcond}) to the integral of the curvature of Berry connection in a
closed loop, which is quantized by the winding number. This necessitates a
'pull back' (i.e., undoing) the lattice transformation of Eq. (\ref{hallcond}%
), i.e., we revert to corresponding matrix-element expressions.

\subsection{'Pull back' of the lattice Weyl transformation}

The pull-back process means we have to undo the lattice transformation of
SFLWT-NEGF transport equation, to return to its \textit{equivalent} matrix
element expressions. Consider the integrand in Eq. (\ref{hallcond}) given by
the partial derivatives of lattice Weyl transformed quantities.%
\begin{align}
&  \left[  \frac{\partial^{\left(  a\right)  }}{\partial\mathcal{\vec{K}}_{x}%
}\frac{\partial^{\left(  b\right)  }}{\partial\mathcal{\vec{K}}_{y}}%
-\frac{\partial^{\left(  a\right)  }}{\partial\mathcal{\vec{K}}_{y}}%
\frac{\partial^{\left(  b\right)  }}{\partial\mathcal{\vec{K}}_{x}}\right]
H^{\left(  a\right)  }\left(  \mathcal{\vec{K}}^{\prime},\mathcal{E}\right)
\left(  -iG^{<\left(  b\right)  }\left(  \mathcal{\vec{K}}^{\prime
},\mathcal{E}\right)  \right) \nonumber\\
&  =\left[  \frac{\partial H^{\left(  a\right)  }\left(  \mathcal{\vec{K}%
}^{\prime},\mathcal{E}\right)  }{\partial\mathcal{\vec{K}}_{x}}\frac{\partial
G^{<\left(  b\right)  }\left(  \mathcal{\vec{K}}^{\prime},\mathcal{E}\right)
}{\partial\mathcal{\vec{K}}_{y}}-\frac{\partial H^{\left(  a\right)  }\left(
\mathcal{\vec{K}}^{\prime},\mathcal{E}\right)  }{\partial\mathcal{\vec{K}}%
_{y}}\frac{\partial G^{<\left(  b\right)  }\left(  \mathcal{\vec{K}}^{\prime
},\mathcal{E}\right)  }{\partial\mathcal{\vec{K}}_{x}}\right]  \text{.}
\label{kxky}%
\end{align}
Take first the term of Eq. (\ref{kxky}), where,%
\begin{equation}
\frac{\partial H^{\left(  a\right)  }\left(  \mathcal{\vec{K}}^{\prime
},\mathcal{E}\right)  }{\partial k_{x}}=\hbar\frac{\partial H^{\left(
a\right)  }\left(  \mathcal{\vec{K}}^{\prime},\mathcal{E}\right)  }%
{\partial\mathcal{\vec{K}}_{x}}\text{.} \label{firsttrm}%
\end{equation}
From Eq. (\ref{sump}) this can be written as a lattice Weyl transform
$\mathcal{W}$ in the form,%
\begin{align}
&  \frac{\partial H^{\left(  a\right)  }\left(  \mathcal{\vec{K}}^{\prime
},\mathcal{E}\right)  }{\partial\mathcal{\vec{K}}_{x}}\nonumber\\
&  =\mathcal{W}\left\{  \left(  \frac{\partial}{\partial\mathcal{\vec{K}}%
_{x}^{\alpha}}+\frac{\partial}{\partial\mathcal{\vec{K}}_{x}^{^{\beta}}%
}\right)  \left\langle \alpha,\mathcal{\vec{K}}^{\prime},\mathcal{E}%
\right\vert \hat{H}\left\vert \beta,\mathcal{\vec{K}}^{\prime},\mathcal{E}%
\right\rangle \right\}  ,\nonumber\\
&  =\mathcal{W}\left\{  \left\langle \alpha,\frac{\partial}{\partial
\mathcal{\vec{K}}_{x}}\mathcal{\vec{K}}^{\prime},\mathcal{E}\right\vert
\hat{H}\left\vert \beta,\mathcal{\vec{K}}^{\prime},\mathcal{E}\right\rangle
+\left\langle \alpha,\mathcal{\vec{K}}^{\prime},\mathcal{E}\right\vert \hat
{H}\left\vert \beta,\frac{\partial}{\partial\mathcal{\vec{K}}_{x}%
}\mathcal{\vec{K}}^{\prime},\mathcal{E}\right\rangle \right\}  ,\nonumber\\
&  =\mathcal{W}\left\{  E_{\beta}\left(  \mathcal{\vec{K}}^{\prime
},\mathcal{E}\right)  \left\langle \alpha,\frac{\partial}{\partial
\mathcal{\vec{K}}_{x}}\mathcal{\vec{K}}^{\prime},\mathcal{E}\right\vert
\left\vert \beta,\mathcal{\vec{K}}^{\prime},\mathcal{E}\right\rangle
+E_{\alpha}\left(  \mathcal{\vec{K}}^{\prime},\mathcal{E}\right)  \left\langle
\alpha,\mathcal{\vec{K}}^{\prime},\mathcal{E}\right\vert \left\vert
\beta,\frac{\partial}{\partial\mathcal{\vec{K}}_{x}}\mathcal{\vec{K}}^{\prime
},\mathcal{E}\right\rangle \right\}  ,\nonumber\\
&  =\mathcal{W}\left\{  \left(  E_{\beta}\left(  \mathcal{\vec{K}}^{\prime
},\mathcal{E}\right)  -E_{\alpha}\left(  \mathcal{\vec{K}}^{\prime
},\mathcal{E}\right)  \right)  \left\langle \alpha,\frac{\partial}%
{\partial\mathcal{\vec{K}}_{x}}\mathcal{\vec{K}}^{\prime},\mathcal{E}%
\right\vert \left\vert \beta,\mathcal{\vec{K}}^{\prime},\mathcal{E}%
\right\rangle \right\}  \text{,} \label{LWTdhdKx}%
\end{align}
where we defined%
\[
\left\langle \alpha,\frac{\partial}{\partial\mathcal{\vec{K}}_{x}%
}\mathcal{\vec{K}}^{\prime},\mathcal{E}\right\vert \equiv\frac{\partial
}{\partial\mathcal{\vec{K}}_{x}}\left\langle \alpha,\mathcal{\vec{K}}^{\prime
},\mathcal{E}\right\vert .
\]

We also have%
\[
\frac{\partial G^{<\left(  b\right)  }\left(  \mathcal{\vec{K}}^{\prime
},\mathcal{E}\right)  }{\partial\mathcal{K}_{y}}=\mathcal{W}\left\{  \left(
\frac{\partial}{\partial\mathcal{\vec{K}}_{y}^{\beta}}+\frac{\partial
}{\partial\mathcal{\vec{K}}_{y}^{^{\alpha}}}\right)  \left\langle
\beta,\mathcal{\vec{K}}^{\prime},\mathcal{E}\right\vert \left(  i\hat{\rho
}\right)  \left\vert \alpha,\mathcal{\vec{K}}^{\prime},\mathcal{E}%
\right\rangle \right\}  \text{,}%
\]
where $\hat{\rho}$ is the density matrix operator. The density operator in the
Heisenberg represntation is given by,
\begin{align}
\hat{\rho}\left(  t\right)   &  =e^{-\frac{i}{\hbar}\hat{H}t}\hat{\rho}\left(
0\right)  e^{\frac{i}{\hbar}\hat{H}t},\nonumber\\
&  =\hat{U}\left(  t\right)  \hat{\rho}\left(  0\right)  \hat{U}^{\dagger
}\left(  t\right)  \text{,} \label{timedepend}%
\end{align}
which provides the major time dependence in our transport equation that
follows. From Eq. (\ref{timedepend}), we take the time dependence of
\[
\left\langle \beta,\mathcal{\vec{K}}^{\prime},\mathcal{E}\right\vert \left(
i\hat{\rho}\right)  \left\vert \alpha,\mathcal{\vec{K}}^{\prime}%
,\mathcal{E}\right\rangle
\]
to be given by
\[
i\left\langle \beta,\mathcal{\vec{K}}^{\prime},\mathcal{E}\right\vert
\hat{\rho}\left(  0\right)  \left\vert \alpha,\mathcal{\vec{K}}^{\prime
},\mathcal{E}\right\rangle e^{i\omega_{\alpha\beta}t}.
\]

We have \ \cite{footnote}
\[
\frac{\partial G^{<\left(  b\right)  }\left(  \mathcal{\vec{K}},\mathcal{E}%
\right)  }{\partial\mathcal{K}_{y}}=\mathcal{W}\left\{
\begin{array}
[c]{c}%
\left\langle \beta,\frac{\partial}{\partial\mathcal{\vec{K}}_{y}}%
\mathcal{\vec{K}}^{\prime},\mathcal{E}\right\vert \left(  i\hat{\rho}%
_{0}\right)  \left\vert \alpha,\mathcal{\vec{K}}^{\prime},\mathcal{E}%
\right\rangle \\
+\left\langle \beta,\mathcal{\vec{K}}^{\prime},\mathcal{E}\right\vert
i\hat{\rho}_{0}\left\vert \alpha,\frac{\partial}{\partial\mathcal{\vec{K}}%
_{y}}\mathcal{\vec{K}}^{\prime},\mathcal{E}\right\rangle
\end{array}
\right\}  e^{i\omega_{\alpha\beta}t}\text{.}%
\]
The density matrix operator $\hat{\rho}_{0}$\ is of the form,%
\begin{align*}
\hat{\rho}_{o}  &  =\sum\limits_{m}\rho_{m}\left\vert m\right\rangle
\left\langle m\right\vert \\
\hat{\rho}_{o}\left\vert m\right\rangle  &  =\rho_{m}\left\vert m\right\rangle
=f\left(  E_{m}\right)  \left\vert m\right\rangle \\
\left\langle m\right\vert \hat{\rho}_{o}\left\vert n\right\rangle  &
=\rho_{mm}=f\left(  E_{n}\right)  \delta_{mn}\text{ or }f\left(  E_{m}\right)
\delta_{mn}%
\end{align*}
where the weight function is the Fermi-Dirac function,%
\[
\rho_{0}^{m}=f\left(  E_{m}\right)
\]
Hence%
\begin{align*}
i\hat{\rho}_{o}\left\vert \alpha,\mathcal{\vec{K}}^{\prime},\mathcal{E}%
\right\rangle  &  =i\sum\limits_{\gamma}\left\vert \gamma,\mathcal{\vec{K}%
}^{\prime},\mathcal{E}\right\rangle \rho_{0}^{\gamma}\left\langle
\gamma,\mathcal{\vec{K}}^{\prime},\mathcal{E}\right\vert \left\vert
\alpha,\mathcal{\vec{K}}^{\prime},\mathcal{E}\right\rangle \\
&  =i\left\vert \alpha,\mathcal{\vec{K}}^{\prime},\mathcal{E}\right\rangle
f\left(  E_{\alpha}\right)  \text{.}%
\end{align*}
Similarly,%
\begin{align*}
i\left\langle \beta,\mathcal{\vec{K}}^{\prime},\mathcal{E}\right\vert \left(
\hat{\rho}_{0}\right)   &  =i\left\langle \beta,\mathcal{\vec{K}}^{\prime
},\mathcal{E}\right\vert \sum\limits_{\gamma}\left\vert \gamma,\mathcal{\vec
{K}}^{\prime},\mathcal{E}\right\rangle \rho_{0}\left\langle \gamma
,\mathcal{\vec{K}}^{\prime},\mathcal{E}\right\vert \\
&  =if\left(  E_{\beta}\right)  \left\langle \beta,\mathcal{\vec{K}}^{\prime
},\mathcal{E}\right\vert \text{.}%
\end{align*}
Hence%
\[
\frac{\partial G^{<\left(  b\right)  }\left(  \mathcal{\vec{K}}^{\prime
},\mathcal{E}\right)  }{\partial\mathcal{K}_{y}}=\mathcal{W}\left\{
\begin{array}
[c]{c}%
i\left\langle \beta,\frac{\partial}{\partial\mathcal{\vec{K}}_{y}%
}\mathcal{\vec{K}}^{\prime},\mathcal{E}\right\vert \left\vert \alpha
,\mathcal{\vec{K}}^{\prime},\mathcal{E}\right\rangle \rho_{0}^{\alpha}\\
i\left\langle \beta,\mathcal{\vec{K}}^{\prime},\mathcal{E}\right\vert
\left\vert \alpha,\frac{\partial}{\partial\mathcal{\vec{K}}_{y}}%
\mathcal{\vec{K}}^{\prime},\mathcal{E}\right\rangle \rho_{0}^{\beta}%
\end{array}
\right\}  e^{i\omega_{\alpha\beta}t}\text{.}%
\]
Shifting the first derivative to the right, we have%
\begin{align*}
\frac{\partial G^{<\left(  b\right)  }\left(  \mathcal{\vec{K}}^{\prime
},\mathcal{E}\right)  }{\partial\mathcal{K}_{y}}  &  =\mathcal{W}\left\{
\begin{array}
[c]{c}%
-i\left\langle \beta,\mathcal{\vec{K}}^{\prime},\mathcal{E}\right\vert
\left\vert \alpha,\frac{\partial}{\partial\mathcal{\vec{K}}_{y}}%
\mathcal{\vec{K}}^{\prime},\mathcal{E}\right\rangle f\left(  E_{\alpha}\right)
\\
i\left\langle \beta,\mathcal{\vec{K}}^{\prime},\mathcal{E}\right\vert
\left\vert \alpha,\frac{\partial}{\partial\mathcal{\vec{K}}_{y}}%
\mathcal{\vec{K}}^{\prime},\mathcal{E}\right\rangle f\left(  E_{\beta}\right)
\end{array}
\right\}  e^{i\omega_{\alpha\beta}t}\\
&  =\mathcal{W}\left[  \left\{  i\left(  f\left(  E_{\beta}\right)  -f\left(
E_{\alpha}\right)  \right)  \left\langle \beta,\mathcal{\vec{K}}%
,\mathcal{E}\right\vert \left\vert \alpha,\frac{\partial}{\partial
\mathcal{\vec{K}}_{y}}\mathcal{\vec{K}}^{\prime},\mathcal{E}\right\rangle
\right\}  e^{i\omega_{\alpha\beta}t}\right]  \text{.}%
\end{align*}
For energy scale it is convenient to chose $f\left(  E_{\alpha}\right)  $ in
the above equation, with the viewpoint that $\alpha$-state, i.e., LLL sub-band
is far remove from the $\beta$-state in gapped states, so that we can set
$f\left(  E_{\beta}\right)  \simeq0$. In the case that several sub-bands are
occupied in FQHE, we can assume that the last occupied sub-band is
fractionally occupied. Thus unoccupied sub-bands will not enter in the
summation. For those sub-bands we can set $f\left(  E_{\beta}\right)  \simeq
0$. Therefore%
\begin{align*}
&  \frac{\partial H\left(  \mathcal{\vec{K}}^{\prime},\mathcal{E}\right)
}{\partial\mathcal{K}_{x}}\frac{\partial G^{<}\left(  \mathcal{\vec{K}%
}^{\prime},\mathcal{E}\right)  }{\partial\mathcal{K}_{y}}\\
&  =\left\{
\begin{array}
[c]{c}%
\mathcal{W}\left[  \left(  E_{\beta}\left(  \mathcal{\vec{K}}^{\prime
},\mathcal{E}\right)  -E_{\alpha}\left(  \mathcal{\vec{K}}^{\prime
},\mathcal{E}\right)  \right)  \left\{  \left\langle \alpha,\frac{\partial
}{\partial\mathcal{\vec{K}}_{x}^{^{^{\prime\prime}}}}\mathcal{\vec{K}}%
^{\prime},\mathcal{E}\right\vert \left\vert \beta,\mathcal{\vec{K}%
;},\mathcal{E}\right\rangle \right\}  \right] \\
\times\mathcal{W}\left[  \left\{  i\left\langle \beta,\mathcal{\vec{K}%
}^{\prime},\mathcal{E}\right\vert \left\vert \alpha,\frac{\partial}%
{\partial\mathcal{\vec{K}}_{y}}\mathcal{\vec{K}}^{\prime},\mathcal{E}%
\right\rangle \right\}  \right]  f\left(  E_{\alpha}\right)  e^{i\omega
_{\alpha\beta}t}%
\end{array}
\right\}  \text{.}%
\end{align*}
Since it appears as a product of two Weyl transforms, it must be a trace
formula in the \textit{untransformed or pulled back} version, i.e., for the
remaing indices $\alpha$ and $\beta$ we must be a summation,
\begin{align*}
&  \frac{\partial H\left(  \mathcal{\vec{K}}^{\prime},\mathcal{E}\right)
}{\partial\mathcal{K}_{x}}\frac{\partial G^{<}\left(  \mathcal{\vec{K}%
}^{\prime},\mathcal{E}\right)  }{\partial\mathcal{K}_{y}}\\
&  =\mathcal{W}\left[
\begin{array}
[c]{c}%
\sum\limits_{\alpha,\beta}\left\{
\begin{array}
[c]{c}%
\left(  E_{\beta}\left(  \mathcal{\vec{K}}^{\prime},\mathcal{E}\right)
-E_{\alpha}\left(  \mathcal{\vec{K}}^{\prime},\mathcal{E}\right)  \right) \\
\times\left\{  \left\langle \alpha,\frac{\partial}{\partial\mathcal{K}_{x}%
}\mathcal{\vec{K}}^{\prime},\mathcal{E}\right\vert \left\vert \beta
,\mathcal{\vec{K}}^{\prime},\mathcal{E}\right\rangle \right\}  \left\{
\left\langle \beta,\mathcal{\vec{K}}^{\prime},\mathcal{E}\right\vert
\left\vert \alpha,\frac{\partial}{\partial\mathcal{K}_{y}}\mathcal{\vec{K}%
}^{\prime},\mathcal{E}\right\rangle \right\}  e^{i\omega_{\alpha\beta}t}%
\end{array}
\right\} \\
\times\ i\left(  f\left(  E_{\alpha}\right)  \right)  .
\end{array}
\right]  \text{.}%
\end{align*}
Similarly, we have%
\begin{align*}
&  \frac{\partial H\left(  \mathcal{\vec{K}}^{\prime},\mathcal{E}\right)
}{\partial\mathcal{K}_{y}}\frac{\partial G^{<}\left(  \mathcal{\vec{K}%
}^{\prime},\mathcal{E}\right)  }{\partial\mathcal{K}_{x}}\\
&  =\mathcal{W}\left[
\begin{array}
[c]{c}%
\sum\limits_{\alpha,\beta}\left\{
\begin{array}
[c]{c}%
\left(  E_{\beta}\left(  \mathcal{\vec{K}}^{\prime},\mathcal{E}\right)
-E_{\alpha}\left(  \mathcal{\vec{K}}^{\prime},\mathcal{E}\right)  \right) \\
\times\left\{  \left\langle \alpha,\frac{\partial}{\partial\mathcal{K}_{y}%
}\mathcal{\vec{K}}^{\prime},\mathcal{E}\right\vert \left\vert \beta
,\mathcal{\vec{K}}^{\prime},\mathcal{E}\right\rangle \right\}  \left\{
\left\langle \beta,\mathcal{\vec{K}}^{\prime},\mathcal{E}\right\vert
\left\vert \alpha,\frac{\partial}{\partial\mathcal{K}_{x}}\mathcal{\vec{K}%
}^{\prime},\mathcal{E}\right\rangle \right\}  e^{i\omega_{\alpha\beta}t}%
\end{array}
\right\} \\
\times\ if\left(  E_{\alpha}\right)  .
\end{array}
\right]  \text{.}%
\end{align*}

Therefore we obtain,%
\begin{align*}
&  \left[  \frac{\partial H\left(  \mathcal{\vec{K}}^{\prime},\mathcal{E}%
\right)  }{\partial\mathcal{K}_{x}}\frac{\partial G^{<}\left(  \mathcal{\vec
{K}}^{\prime},\mathcal{E}\right)  }{\partial\mathcal{K}_{y}}-\frac{\partial
H\left(  \mathcal{\vec{K}}^{\prime},\mathcal{E}\right)  }{\partial
\mathcal{K}_{y}}\frac{\partial G^{<}\left(  \mathcal{\vec{K}}^{\prime
},\mathcal{E}\right)  }{\partial\mathcal{K}_{x}}\right] \\
&  =\mathcal{W}\left[
\begin{array}
[c]{c}%
\sum\limits_{\alpha,\beta}\left\{
\begin{array}
[c]{c}%
\left(  E_{\beta}\left(  \mathcal{\vec{K}}^{\prime},\mathcal{E}\right)
-E_{\alpha}\left(  \mathcal{\vec{K}}^{\prime},\mathcal{E}\right)  \right) \\
\times\left[
\begin{array}
[c]{c}%
\left\{  \left\langle \alpha,\frac{\partial}{\partial\mathcal{K}_{x}%
}\mathcal{\vec{K}}^{\prime},\mathcal{E}\right\vert \left\vert \beta
,\mathcal{\vec{K}}^{\prime},\mathcal{E}\right\rangle \right\}  \left\{
\left\langle \beta,\mathcal{\vec{K}}^{\prime},\mathcal{E}\right\vert
\left\vert \alpha,\frac{\partial}{\partial\mathcal{K}_{y}}\mathcal{\vec{K}%
}^{\prime},\mathcal{E}\right\rangle \right\} \\
-\left\{  \left\langle \alpha,\frac{\partial}{\partial\mathcal{K}_{y}%
}\mathcal{\vec{K}}^{\prime},\mathcal{E}\right\vert \left\vert \beta
,\mathcal{\vec{K}}^{\prime},\mathcal{E}\right\rangle \right\}  \left\{
\left\langle \beta,\mathcal{\vec{K}}^{\prime},\mathcal{E}\right\vert
\left\vert \alpha,\frac{\partial}{\partial\mathcal{K}_{x}}\mathcal{\vec{K}%
}^{\prime},\mathcal{E}\right\rangle \right\}
\end{array}
\right]
\end{array}
\right\} \\
\times\ ie^{i\omega_{\alpha\beta}t}\left(  f\left(  E_{\alpha}\right)
\right)
\end{array}
\right]  \text{.}%
\end{align*}
Now the LHS of Eq. (\ref{eqKE}), namely%
\begin{align}
&  \left(  \frac{a}{\left(  2\pi\hbar\right)  }\right)  ^{2}\int dp_{x}%
dp_{y}\frac{e}{a^{2}}\frac{\partial\mathcal{E}}{\partial\mathcal{\vec{K}}_{y}%
}G^{<}\left(  \mathcal{\vec{K}}^{\prime},\mathcal{E}\right) \nonumber\\
&  =\left(  \frac{a}{\left(  2\pi\hbar\right)  }\right)  ^{2}\int dp_{x}%
dp_{y}\frac{e}{a^{2}}\frac{\partial H}{\partial p_{y}}G^{<}\left(
\mathcal{\vec{K}}^{\prime},\mathcal{E}\right)  \text{.} \label{LHS}%
\end{align}
Using the result of Eq. (\ref{LWTdhdKx}), we have%
\begin{align*}
&  \frac{\partial H\left(  \mathcal{\vec{K}}^{\prime},\mathcal{E}\right)
}{\partial\mathcal{\vec{K}}_{y}}\\
&  =\mathcal{W}\left\{  \left[  \left(  E_{\alpha}\left(  \mathcal{\vec{K}%
}^{\prime},\mathcal{E}\right)  -E_{\beta}\left(  \mathcal{\vec{K}}^{\prime
},\mathcal{E}\right)  \right)  \right]  \left\langle \alpha,\mathcal{\vec{K}%
}^{\prime},\mathcal{E}\right\vert \left\vert \beta,\frac{\partial}%
{\partial\mathcal{\vec{K}}_{y}}\mathcal{\vec{K}}^{\prime},\mathcal{E}%
\right\rangle \right\}  ,\\
&  =\mathcal{W}\left\{  \left[  \left(  E_{\beta}\left(  \mathcal{\vec{K}%
}^{\prime},\mathcal{E}\right)  -E_{\alpha}\left(  \mathcal{\vec{K}}^{\prime
},\mathcal{E}\right)  \right)  \right]  \left\langle \alpha,\frac{\partial
}{\partial\mathcal{\vec{K}}_{y}}\mathcal{\vec{K}}^{\prime},\mathcal{E}%
\right\vert \left\vert \beta,\mathcal{\vec{K}}^{\prime},\mathcal{E}%
\right\rangle \right\}  ,\\
&  =\mathcal{W}\left\{  \omega_{\beta\alpha}\left\langle \alpha,\frac
{\partial}{\partial k_{y}}\mathcal{\vec{K}}^{\prime},\mathcal{E}\right\vert
\left\vert \beta,\mathcal{\vec{K}}^{\prime},\mathcal{E}\right\rangle \right\}
=\left\langle \alpha,\mathcal{\vec{K}}^{\prime},\mathcal{E}\right\vert
v_{g,y}\left\vert \beta,\mathcal{\vec{K}}^{\prime},\mathcal{E}\right\rangle ,
\end{align*}
where we use the identity%
\[
\omega_{\beta\alpha}\left\langle \alpha,\nabla_{\vec{p}}\vec{p}\right\vert
\left\vert \beta,\vec{p}\right\rangle =\left\langle \alpha,\vec{p}\right\vert
\vec{v}_{g}\left\vert \beta,\vec{p}\right\rangle
\]
Likewise%
\[
G^{<}\left(  \mathcal{\vec{K}},\mathcal{E}\right)  =i\mathcal{W}\left(
\left\langle \beta,\mathcal{\vec{K}}^{\prime},\mathcal{E}\right\vert \hat
{\rho}\left\vert \alpha,\mathcal{\vec{K}}^{\prime},\mathcal{E}\right\rangle
\right)
\]
Again, since Eq. (\ref{LHS}) is a product of lattice Weyl transform, it must
be a trace in the \textit{untransformed} version, i.e.,
\begin{align*}
&  \left(  \frac{a}{\left(  2\pi\hbar\right)  }\right)  ^{2}\int dp_{x}%
dp_{y}\frac{e}{a^{2}}\frac{\partial H}{\partial\mathcal{\vec{K}}_{y}}%
G^{<}\left(  \mathcal{\vec{K}},\mathcal{E}\right) \\
&  =\mathcal{W}\left\{
\begin{array}
[c]{c}%
i\int\left(  \frac{a}{\left(  2\pi\hbar\right)  }\right)  ^{2}dp_{x}dp_{y}\\
\times\sum\limits_{\alpha,\beta}\left\langle \alpha,\mathcal{\vec{K}}^{\prime
},\mathcal{E}\right\vert \frac{e}{a^{2}}v_{y}\left\vert \beta,\mathcal{\vec
{K}}^{\prime},\mathcal{E}\right\rangle \left\langle \beta,\mathcal{\vec{K}%
}^{\prime},\mathcal{E}\right\vert \hat{\rho}\left\vert \alpha,\mathcal{\vec
{K}}^{\prime},\mathcal{E}\right\rangle
\end{array}
\right\}  ,\\
&  =\mathcal{W}\left\{  iTr\left(  \frac{e}{a^{2}}\hat{v}_{g,y}\right)
\hat{\rho}\right\}  =i\mathcal{W}\left\{  Tr\left(  \hat{\jmath}_{y}\hat{\rho
}\right)  \right\}  =i\mathcal{W}\left\{  Tr\left(  \hat{\jmath}_{y}%
\ \hat{\rho}\right)  \right\}  ,\\
&  =i\mathcal{W}\left\{  \left\langle \hat{\jmath}_{y}\left(  t\right)
\right\rangle \right\}  \text{.}%
\end{align*}
For calculating the conductivity we are interested in the term multiplying the
first-order in electric field. We can now convert the quantum transport
equation in the transformed space, Eq. (\ref{eqKE}),
\begin{align*}
&  \left(  \frac{a}{\left(  2\pi\hbar\right)  }\right)  ^{2}\int dp_{x}%
dp_{y}\frac{e}{a^{2}}\frac{\partial\mathcal{E}}{\partial\mathcal{\vec{K}}%
_{y}^{\prime}}\left[  -iG^{<}\left(  \mathcal{\vec{K}}^{\prime},\mathcal{E}%
\right)  \right] \\
&  =\left(  \frac{1}{k}\right)  e^{2}\left\vert \vec{F}\right\vert \frac
{1}{\left(  2\pi\hbar\right)  ^{2}}\int\int\int dp_{x}dp_{y}dt\ \left[
\frac{\partial^{\left(  a\right)  }}{\partial\mathcal{\vec{K}}_{x}}%
\frac{\partial^{\left(  b\right)  }}{\partial\mathcal{\vec{K}}_{y}}%
-\frac{\partial^{\left(  a\right)  }}{\partial\mathcal{\vec{K}}_{y}}%
\frac{\partial^{\left(  b\right)  }}{\partial\mathcal{\vec{K}}_{x}}\right] \\
&  \times H^{\left(  a\right)  }\left(  \mathcal{\vec{K}}^{\prime}%
,\mathcal{E}\right)  \left(  -iG^{<\left(  b\right)  }\left(  \mathcal{\vec
{K}}^{\prime},\mathcal{E}\right)  \right)  \text{,}%
\end{align*}
to the equivalent matrix element expressions by undoing the lattice Weyl
transformation $\mathcal{W}$, which amounts to canceling $\mathcal{W}$ in both
side of the equation given by,
\begin{align}
&  \mathcal{W}\left\{  \left\langle \hat{\jmath}_{y}\left(  \bar{t}\right)
\right\rangle \right\} \nonumber\\
&  =\mathcal{W}\left\{
\begin{array}
[c]{c}%
\left(  \frac{1}{k}\right)  \frac{e^{2}}{h}\left\vert \vec{F}\right\vert
\frac{1}{\left(  2\pi\hbar\right)  }\int\int\int dp_{x}dp_{y}dt\\
\times\sum\limits_{\alpha,\beta}\left[
\begin{array}
[c]{c}%
\left(  E_{\beta}\left(  \mathcal{\vec{K}}^{\prime},\mathcal{E}\right)
-E_{\alpha}\left(  \mathcal{\vec{K}}^{\prime},\mathcal{E}\right)  \right) \\
\times\left\{
\begin{array}
[c]{c}%
\left\langle \alpha,\frac{\partial}{\partial\mathcal{\vec{K}}_{x}%
}\mathcal{\vec{K}}^{\prime},\mathcal{E}\right\vert \left\vert \beta
,\mathcal{\vec{K}}^{\prime},\mathcal{E}\right\rangle \left\langle
\beta,\mathcal{\vec{K}}^{\prime},\mathcal{E}\right\vert \left\vert
\alpha,\frac{\partial}{\partial\mathcal{\vec{K}}_{y}}\mathcal{\vec{K}}%
^{\prime},\mathcal{E}\right\rangle \\
-\left\langle \alpha,\frac{\partial}{\partial\mathcal{\vec{K}}_{y}%
}\mathcal{\vec{K}}^{\prime},\mathcal{E}\right\vert \left\vert \beta
,\mathcal{\vec{K}}^{\prime},\mathcal{E}\right\rangle \left\langle
\beta,\mathcal{\vec{K}}^{\prime},\mathcal{E}\right\vert \left\vert
\alpha,\frac{\partial}{\partial\mathcal{\vec{K}}_{x}}\mathcal{\vec{K}}%
^{\prime},\mathcal{E}\right\rangle
\end{array}
\right\} \\
\times\ e^{i\omega_{\alpha\beta}t}f\left(  E_{\alpha}\right)
\end{array}
\right]  \text{.}%
\end{array}
\right\} \nonumber\\
&  \label{pullback}%
\end{align}
The time integral of the RHS amounts to taking zero-order time dependence
[zero electric field] of the rest of the integrand, then we have for the
remaining time-dependence, explicitly integrated as,
\begin{align*}
\int\limits_{-\infty}^{0}dt\exp i\omega_{\alpha\beta}t  &  =\left.  \frac
{\exp\exp i\omega_{\alpha\beta}t}{i\omega_{\alpha\beta}}\right\vert
_{\tau=-\infty}^{\tau=0}\\
&  =\left.  \frac{\exp\left(  i\left(  \omega_{\alpha\beta}-i\eta\right)
\tau\right)  }{i\omega_{\alpha\beta}}\right\vert _{\tau=-\infty}^{\tau
=0}=\frac{1}{i\omega_{\alpha\beta}}%
\end{align*}
Thus eliminating the time integral we finally obtain.
\begin{align*}
&  \left\langle \hat{\jmath}_{y}\left(  t\right)  \right\rangle \\
&  =-i\frac{e^{2}}{h}\left(  \frac{1}{k}\right)  \left\vert \vec{F}\right\vert
\frac{1}{\left(  2\pi\hbar\right)  }\int\int dp_{x}dp_{y}\\
&  \times\sum\limits_{\alpha,\beta}\left[
\begin{array}
[c]{c}%
f\left(  E_{\alpha}\right)  \left(  -\frac{\hbar\omega_{\alpha\beta}}%
{\omega_{\alpha\beta}}\right) \\
\times\left\{
\begin{array}
[c]{c}%
\left\langle \alpha,\frac{\partial}{\partial\mathcal{\vec{K}}_{x}%
}\mathcal{\vec{K}}^{\prime},\mathcal{E}\right\vert \left\vert \beta
,\mathcal{\vec{K}}^{\prime},\mathcal{E}\right\rangle \left\langle
\beta,\mathcal{\vec{K}}^{\prime},\mathcal{E}\right\vert \left\vert
\alpha,\frac{\partial}{\partial\mathcal{\vec{K}}_{y}}\mathcal{\vec{K}}%
^{\prime},\mathcal{E}\right\rangle \\
-\left\langle \alpha,\frac{\partial}{\partial\mathcal{\vec{K}}_{y}%
}\mathcal{\vec{K}}^{\prime},\mathcal{E}\right\vert \left\vert \beta
,\mathcal{\vec{K}}^{\prime},\mathcal{E}\right\rangle \left\langle
\beta,\mathcal{\vec{K}}^{\prime},\mathcal{E}\right\vert \left\vert
\alpha,\frac{\partial}{\partial\mathcal{\vec{K}}_{x}}\mathcal{\vec{K}}%
^{\prime},\mathcal{E}\right\rangle
\end{array}
\right\}
\end{array}
\right]
\end{align*}
which reduces to%
\begin{align}
&  \left\langle \hat{\jmath}_{y}\left(  t\right)  \right\rangle \nonumber\\
&  =\left(  \frac{1}{k}\right)  i\frac{e^{2}}{h}\left\vert \vec{F}\right\vert
\frac{1}{\left(  2\pi\right)  }\int\int dk_{x}dk_{y}\nonumber\\
&  \times\sum\limits_{\alpha,\beta}f\left(  E_{\alpha}\right)  \left\{
\begin{array}
[c]{c}%
\left\langle \alpha,\frac{\partial}{\partial k_{x}}\mathcal{\vec{K}}^{\prime
},\mathcal{E}\right\vert \left\vert \beta,\mathcal{\vec{K}}^{\prime
},\mathcal{E}\right\rangle \left\langle \beta,\mathcal{\vec{K}}^{\prime
},\mathcal{E}\right\vert \left\vert \alpha,\frac{\partial}{\partial k_{y}%
}\mathcal{\vec{K}}^{\prime},\mathcal{E}\right\rangle \\
-\left\langle \alpha,\frac{\partial}{\partial k_{y}}\mathcal{\vec{K}}^{\prime
},\mathcal{E}\right\vert \left\vert \beta,\mathcal{\vec{K}}^{\prime
},\mathcal{E}\right\rangle \left\langle \beta,\mathcal{\vec{K}}^{\prime
},\mathcal{E}\right\vert \left\vert \alpha,\frac{\partial}{\partial k_{x}%
}\mathcal{\vec{K}}^{\prime},\mathcal{E}\right\rangle
\end{array}
\right\}  \label{notint}%
\end{align}
Taking the Fourier transform of both sides, we obtain%
\begin{align}
&  \left\langle \hat{\jmath}_{y}\left(  \omega\right)  \right\rangle
\nonumber\\
&  =\left(  \frac{1}{k}\right)  i\frac{e^{2}}{h}\left\vert \vec{F}\right\vert
\frac{\delta\left(  \omega\right)  }{\left(  2\pi\right)  }\int\int
dk_{x}dk_{y}\nonumber\\
&  \times\sum\limits_{\alpha,\beta}\left\{
\begin{array}
[c]{c}%
\left\langle \alpha,\frac{\partial}{\partial k_{x}}\mathcal{\vec{K}}^{\prime
},\mathcal{E}\right\vert \left\vert \beta,\mathcal{\vec{K}}^{\prime
},\mathcal{E}\right\rangle \left\langle \beta,\mathcal{\vec{K}}^{\prime
},\mathcal{E}\right\vert \left\vert \alpha,\frac{\partial}{\partial k_{y}%
}\mathcal{\vec{K}}^{\prime},\mathcal{E}\right\rangle \\
-\left\langle \alpha,\frac{\partial}{\partial k_{y}}\mathcal{\vec{K}}^{\prime
},\mathcal{E}\right\vert \left\vert \beta,\mathcal{\vec{K}}^{\prime
},\mathcal{E}\right\rangle \left\langle \beta,\mathcal{\vec{K}}^{\prime
},\mathcal{E}\right\vert \left\vert \alpha,\frac{\partial}{\partial k_{x}%
}\mathcal{\vec{K}}^{\prime},\mathcal{E}\right\rangle
\end{array}
\right\}  f\left(  E_{\alpha}\right)  . \label{currenteq}%
\end{align}
Taking the limit $\omega\Longrightarrow0$ and summing over the states $\beta$,
we readily obtain the conductivity, $\sigma_{yx}$.
\begin{equation}
\sigma_{yx}=\frac{e^{2}}{h}\left(  \frac{1}{k}\right)  \sum\limits_{\alpha
}f\left(  E_{\alpha}\right)  \frac{i}{\left(  2\pi\right)  }\int\int
dk_{x}dk_{y}\left[
\begin{array}
[c]{c}%
\left\langle \alpha,\frac{\partial}{\partial k_{x}}\mathcal{\vec{K}}^{\prime
},\mathcal{E}\right\vert \left\vert \alpha,\frac{\partial}{\partial k_{y}%
}\mathcal{\vec{K}}^{\prime},\mathcal{E}\right\rangle \\
-\left\langle \alpha,\frac{\partial}{\partial k_{y}}\mathcal{\vec{K}}^{\prime
},\mathcal{E}\right\vert \left\vert \alpha,\frac{\partial}{\partial k_{x}%
}\mathcal{\vec{K}}^{\prime},\mathcal{E}\right\rangle
\end{array}
\right]  \text{.} \label{hall}%
\end{equation}
Note the transformation from $\mathcal{\vec{K}}$ $\Longrightarrow$ $\vec{p}$
in the integration in both Eqs. (\ref{currenteq}) and (\ref{hall}) has a
Jacobian unity. Equation (\ref{hall}) is the same expression that can obtained
to derive the integer quantum Hall effect from Kubo formula \cite{previous}.

We now prove that for each statevector, $\left\vert \alpha,\vec{k}%
\right\rangle $, the expression,
\begin{equation}
\left(  \frac{1}{k}\right)  \frac{i}{\left(  2\pi\right)  }\int\int
dk_{x}dk_{y}\ f\left(  E_{\alpha}\left(  \vec{k}\right)  \right)  \left[
\left\langle \frac{\partial}{\partial k_{x}}\alpha,\vec{k}\right\vert
\frac{\partial}{\partial k_{y}}\left\vert \alpha,\vec{k}\right\rangle
-\left\langle \frac{\partial}{\partial k_{y}}\alpha,\vec{k}\right\vert
\frac{\partial}{\partial k_{x}}\left\vert \alpha,\vec{k}\right\rangle \right]
\text{,} \label{chernNum}%
\end{equation}
is the winding number around the occupied contour in the Brillouin zone. First
we can rewrite the terms within the square bracket as%
\begin{align}
&  \left[  \left\langle \frac{\partial}{\partial k_{x}}\alpha,\vec
{k}\right\vert \frac{\partial}{\partial k_{y}}\left\vert \alpha,\vec
{k}\right\rangle -\left\langle \frac{\partial}{\partial k_{y}}\alpha,\vec
{k}\right\vert \frac{\partial}{\partial k_{x}}\left\vert \alpha,\vec
{k}\right\rangle \right] \nonumber\\
&  =\left\langle \frac{\partial}{\partial\vec{k}}\alpha,\vec{k}\right\vert
\times\frac{\partial}{\partial\vec{k}}\left\vert \alpha,\vec{k}\right\rangle
=\nabla_{\vec{k}}\times\left\langle \alpha,\vec{k}\right\vert \frac{\partial
}{\partial\vec{k}}\left\vert \alpha,\vec{k}\right\rangle \text{.}
\label{curlA}%
\end{align}
The last term indicates the operation of the curl of the Berry connection
which is related to the quantization of Hall conductivity. This quantization
is due to the uniqueness of the parallel-transported wavefunction, which may
also have bearing on the self-consistent Bohr-Sommerfeld quantization
\cite{fego}.

At low temperature, we can just write Eq. (\ref{hall}) as,%
\begin{align}
\ \sigma_{yx}  &  =\left(  \frac{1}{k}\right)  \frac{ie^{2}}{2\pi\hbar}%
\frac{1}{\left(  2\pi\right)  }\int\int_{sub-band}dk_{x}dk_{y}\ \left[
\nabla_{\vec{k}}\times\left\langle \alpha,\vec{k}\right\vert \frac{\partial
}{\partial\vec{k}}\left\vert \alpha,\vec{k}\right\rangle \right]
_{plane},\nonumber\\
&  =\left(  \frac{1}{k}\right)  \frac{e^{2}}{2\pi\hbar}\left(  \frac
{1}{\left(  2\pi\right)  }%
{\displaystyle\oint_{sub-band}}
dk_{c}\ \left[  \left\langle LL,\vec{k}\right\vert i\frac{\partial}{\partial
k_{c}}\left\vert LL,\vec{k}\right\rangle \right]  _{contour}\right)  \text{.}
\label{lowT}%
\end{align}
where the quantity within the parenthesis gives $n_{LL}\ \epsilon\
\mathbb{Z}
$ which is the winding number or the Chern number. In the experiment,
$n_{LL}=1$ Therefore%
\begin{align}
\ \sigma_{yx}  &  =\left(  \frac{1}{k}\right)  \frac{e^{2}}{h}\text{,}%
\nonumber\\
&  =\frac{e^{2}}{h}v \label{FQHE3}%
\end{align}

For lower magnetic fields below the magnetic field strength of the fully
occupied LLL, the lowered fully occupied sub-bands simply serve a a background
for the Hall conductance, where the last fractionally occupied sub-band
dictates the FQHE. Thus the FQHE conductivity is given by $\sigma_{yx}%
=\frac{e^{2}}{h}\nu$ below and above the magnetic fields of the fully occupied
LLL. This is what is found in the experiments for magnetic field strength
beyond and below the fully occupied LLL. We emphasize that the prime-number
remainders in going from $v=1$ to $v=2$ in the experiments increases as the
magnetic fields decreases.


\begin{thebibliography}{99}
\bibitem {footnote}Here we use the definition of Green's function without the
factor $\hbar$, following traditional treatments, i.e. $\rho\left(
1,2\right)  =-iG^{<}\left(  1,2\right)  $.
\end{thebibliography}
\end{document}